\documentclass[aps,prl,reprint,amsmath,amssymb,superscriptaddress, 10pt]{revtex4-2}
\usepackage[usenames,dvipsnames]{xcolor}
\usepackage{amsmath,amssymb,amsfonts,latexsym}\usepackage{amsmath,bm}
\usepackage[margin=12mm,includehead,includefoot]{geometry}
\usepackage{graphicx}
\usepackage{braket}
\usepackage[normalem]{ulem}
\usepackage{bbding}
\usepackage{nicefrac}
\usepackage{lipsum}
\usepackage{mathrsfs}
\usepackage{physics}
\usepackage[protrusion=true,expansion=true]{microtype}
\usepackage{times}
\usepackage{hyperref}
\hypersetup{
  pdftitle={Interacting Quantum Symmetric Exclusion Process},
  pdfauthor={Denis Bernard, Friedrich Hübner, Stefano Scopa},
  pdfsubject={},
  pdfkeywords={
    quantum symmetric exclusion process,
    symmetric exclusion process,
    facilitated exclusion process,
    hydrodynamic limit,
    large deviations,
    scaling theory,
    noisy spin chains,
    stochastic quantum dynamics,
    macroscopic fluctuation theory,
    quantum transport,
    diffusive systems,
    coherence fluctuations,
    kinetically constrained models
  },
    colorlinks=true,
    citecolor=blue,
    linkcolor=blue,
    urlcolor=blue
  }
\newcommand{\I}{\text{i}}
\newcommand{\E}{{\mathbb{E}}}

\newcommand{\nn}{\nonumber\\[4pt]}
\newcommand{\Pt}{\hat{p}}
\newcommand{\G}{{\mathcal{G}}}
\newcommand{\Cum}{\kappa}
\newcommand{\Pnot}[1]{\hat P^{(\neg #1)}}
\newcommand{\Pyes}[1]{\hat P^{(#1)}}
\newcommand{\circled}[1]{\tikz[baseline=(char.base)]{\node[shape=circle,draw=blue,text=blue,inner sep=0.2pt,minimum size=1.1em,font=\scriptsize] (char) {#1};}}
\newcommand{\rcircled}[1]{\tikz[baseline=(char.base)]{\node[shape=circle,draw=red,text=red,inner sep=0.2pt,minimum size=1.1em,font=\scriptsize] (char) {#1};}}
\newcommand{\avg}[1]{\langle\!\langle #1 \rangle\!\rangle}
\newcommand{\be}{\begin{equation}}
\newcommand{\ee}{\end{equation}}
\newcommand{\de}{\partial}
\def\bes{\begin{subequations}}
\def\esu{\end{subequations}}
\definecolor{red}{rgb}{0.8,0,0.15}
\definecolor{blue}{rgb}{0.15, 0.15, .8}
\definecolor{green}{rgb}{0., 0.42, .24}
\usepackage{etoolbox}
\usepackage{orcidlink}
\newcommand{\prlsection}[1]{ \noindent\emph{#1---}}

\begin{document}

\newcommand{\titleinfo}{Interacting Quantum Symmetric Exclusion Process}

\title{\titleinfo}

\author{Denis Bernard~\orcidlink{0000-0002-5492-8360}}
\email{denis.bernard@phys.ens.fr}
\affiliation{Laboratoire de Physique de l’\'Ecole Normale Superieure, CNRS,
ENS \& Universit\'e PSL, Sorbonne Universit\'e, Universit\'e Paris Cit\'e, 75005 Paris, France.}

\author{Friedrich H\"ubner~\orcidlink{0000-0003-0932-5322}}
\affiliation{Laboratoire de Physique de l’\'Ecole Normale Superieure, CNRS,
ENS \& Universit\'e PSL, Sorbonne Universit\'e, Universit\'e Paris Cit\'e, 75005 Paris, France.}

\author{Stefano Scopa~\orcidlink{0000-0001-7638-8804}}
\email{stefano.scopa@phys.ens.fr}
\affiliation{Laboratoire de Physique de l’\'Ecole Normale Superieure, CNRS,
ENS \& Universit\'e PSL, Sorbonne Universit\'e, Universit\'e Paris Cit\'e, 75005 Paris, France.}

\begin{abstract}
We introduce and solve the Interacting Quantum Symmetric Exclusion Process (IQSEP), a family of models describing the stochastic quantum hopping of charged particles along the edges of a lattice, with hopping amplitudes that depend on the occupations of neighbouring sites. In the absence of interactions, they reduce to the standard quantum simple symmetric exclusion process, exhibiting coherent diffusive transport. For interactions of order one, they capture incoherent diffusive transport and its fluctuations, characterized by density-dependent diffusivity and mobility, making contact with the macroscopic fluctuation theory.
By rescaling the interaction strength appropriately with the lattice mesh, we define a mesoscopic scaling regime that retains a finite coherence length in the continuous thermodynamic limit. This regime interpolates between coherent behavior at small length scales and incoherent behavior at large scales. The resulting scaling theory accounts for fluctuations of quantum coherences in interacting diffusive systems, going beyond the scope of standard fluctuating hydrodynamics. 
\end{abstract}

\maketitle

\prlsection{Introduction.} 
Understanding the large-scale dynamics of noisy many-body systems is a long-standing and fascinating topic in statistical physics~\cite{Liggett1985,Spohn1991}. At late times, microscopic details enter only through a few transport coefficients, such as diffusion constants and mobilities, and a universal description is expected to emerge~\cite{Spohn1991,Kipnis1999,Kipnis1989}. This viewpoint is particularly powerful in classical systems, where the Macroscopic Fluctuation Theory (MFT)~\cite{bertini2001fluctuations,bertini2002macroscopic,Bertini2005,Bertini2015} provides a precise description not only of typical diffusive profiles, but also of their fluctuations and large deviations~\cite{Derrida2001,Derrida2007,Derrida2011,Mallick2015}. In this context, a timely but more delicate question is whether an analogous universal theory can be formulated for quantum many-body systems~\cite{Bernard2021}. Such a theory should go beyond the framework of fluctuating hydrodynamics, such as the MFT, by coding for fluctuations of quantum coherent phenomena in the mesoscopic regime.

A concrete step in this direction was provided by charge-conserving random-unitary circuits~\cite{nahum2018operator,khemani2018operator,vonKeyserlingk2018operator,diffusive2018rakovszky,Rakovszky2019,Gullans2019,FCSMFTDeNardis} and, in parallel, by continuous-time stochastic formulations such as the Quantum Symmetric Simple Exclusion Process (QSSEP)~\cite{Bauer2017,Bauer2019, Bernard2018,Bernard2019,Bernard2023, Hruza2023,Bernard2024,Bauer2024,Bernard2025,albert2026,Bernard2026-qssep-cont,Barraquand2026}. In QSSEP, fermions hop between neighbouring sites with Brownian amplitudes. The stochastic Hamiltonian is quadratic, so that each noise realization preserves the Gaussian structure of the state. After noise averaging, the dynamics reduces to that of a classical simple exclusion process~\cite{Derrida2007, Mallick2015}. Nevertheless, the stochastic quantum state contains more information than its averaged counterpart. The multi-replica theory carries nontrivial information on coherences and their fluctuations, encoded in so-called coherence loops, which survive at large scales. Interestingly, these coherence loops fully encode the MFT fluctuation statistics of densities and currents~\cite{Bauer2024,Bernard2025,costa2025,albert2026}, but also additional quantum effects that are absent at the classical level~\cite{albert2026}. This makes QSSEP, together with its related
extensions~\cite{alba2025nuqssep, russotto2026, russotto2026b, Minoguchi2023}, a rare analytically tractable setting in which fluctuations of diffusive transport and quantum coherent effects can be studied within the same microscopic framework.

The next natural question is what remains of this structure in the presence of interactions. Interactions are expected to modify transport, possibly producing density-dependent diffusion and nonlinear fluctuating hydrodynamics e.g.~\cite{Spohn1991,Funaki1991,Gonalves2009,Blondel2020,Erignoux2024}. At the same time, they are expected to dephase quantum coherences~\cite{McCulloch2026,McCulloch2026b}. These two effects need not occur at the same scale. The single-replica sector is the one in which MFT is expected to emerge, while multi-replica observables carry the quantum coherent effects~\cite{Bernard2019,Bernard2021,Hruza2023}. A controlled interacting model should therefore allow one to compute both sectors and to identify the scale at which the crossover from coherent to incoherent diffusion takes place.

\begin{figure}[t]
\centering
(a)~\includegraphics[width=.75\columnwidth]{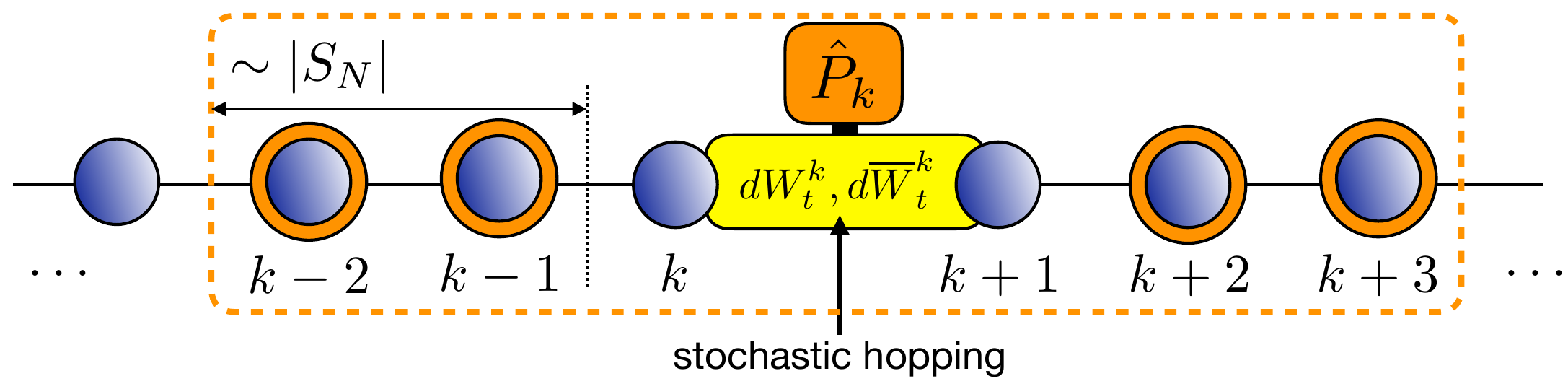}\\[5pt]
(b)~\includegraphics[width=.75\columnwidth]{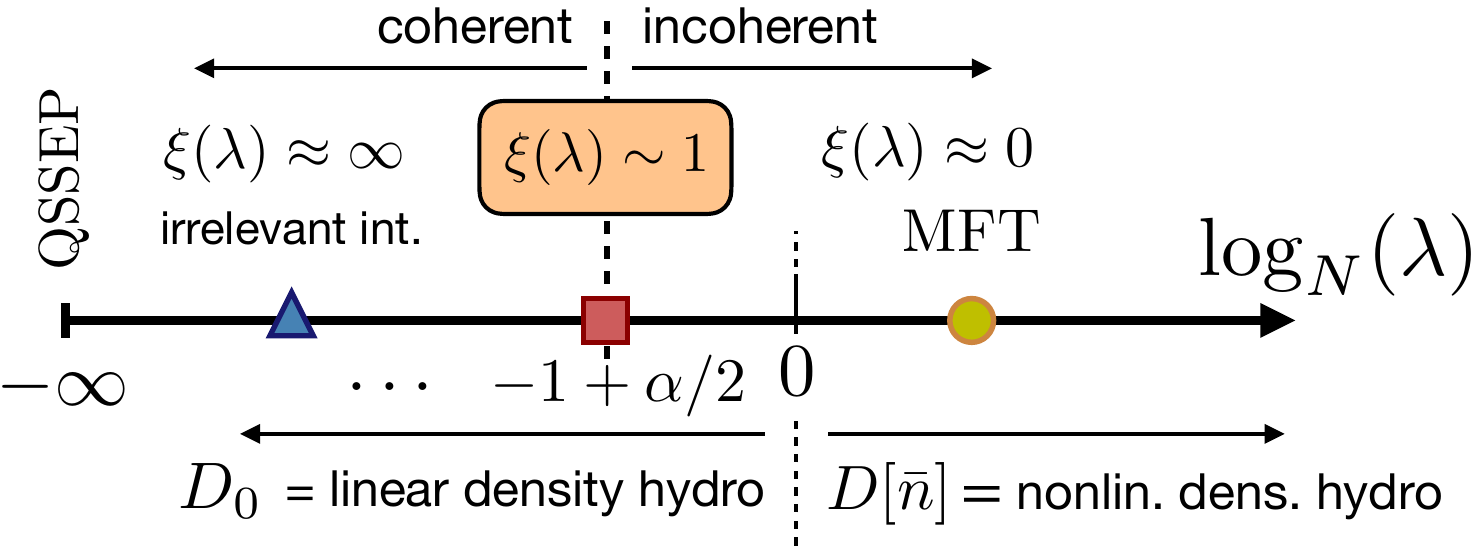}
\caption{(a)~Illustration of the IQSEP model: stochastic nearest-neighbor hopping is dressed by interactions with the occupations of sites within a surrounding subextensive region.  (b)~Phase diagram of the IQSEP model discussed in the main text. The symbols indicate the regimes tested numerically in Fig.~\ref{fig:num}.}\label{fig:illustration}
\end{figure}

In this Letter, we introduce a family of such models: an interacting deformation of QSSEP, which we call IQSEP. This interacting model adds a kinetic dressing to the QSSEP dynamics, by making the stochastic hopping across a link of the chain depend on the occupations of neighbouring sites. This dressing may be viewed as an effective description of inter-particle multiple scattering at ultra-short time and length scales~\cite{Hruza2023}. IQSEP therefore belongs to the broad class of kinetically constrained models, including, for instance, facilitated classical exclusion processes~\cite{Funaki1991, Kob1993,Rossi2000, deOliveira2005, Gonalves2009, Gabel2010, Baik2018, Goldstein2019, Blondel2020, Ayyer2023,Barraquand2025, DaCunha2026} and, on the quantum side, East-type models~\cite{Olmos2012,Olmos2014,Pancotti2020,Rose2022,Brighi2023,Bertini2024} or folded XXZ chains~\cite{Zadnik2021,Zadnik2021b,Pozsgay2021}. This choice is also motivated by earlier works on QSSEP~\cite{Bauer2017}, where the slow dynamics of the dephased XXZ chain is reduced to Brownian hopping constrained by the occupations of neighbouring sites.\\
\indent
 More precisely, we shall consider a one-dimensional chain of size $N+1$, and focus on the diffusive scaling limit $N\to\infty$, with rescaled position $x=i/N$ and time $\tau=t/N^2$ fixed. The stochastic hopping on the link $(k,k+1)$ is dressed by a symmetric density-dependent operator, coding for the impact of the local density environment,
\be\label{eq:P-dressing}
\hat P_k=\sqrt{D_0}\bigl(1+\lambda \Pt_k\bigr),
\ee
with $\Pt_k\equiv p(\{\hat n_{k+q}\}_{q\in S_N})$ a generic function of the occupations, and $D_0={\cal O}(1)$ is the bare diffusion constant. Here $\lambda$ controls the interaction strength,
while $S_N$ denotes the support of $\Pt_k$. We take this support to be subextensive, $|S_N|={\cal O}(N^\alpha)$ with $0\leq\alpha<1$, the case $\alpha=0$ corresponding to a finite-range dressing. An illustration of the IQSEP model is shown in Fig.~\ref{fig:illustration}(a).

\indent
As discussed below, the effect of the interactions is encoded in the emergence of a tunable coherence length $\xi(\lambda)\sim \sqrt{l_\text{uv} l_\text{int}}/\lambda$, with $l_\text{uv}\sim a_0/N$ the microscopic scale, and $l_\text{int}\sim a_0|S_N|/N$ the mesoscopic scale sets by the interaction; $a_0\equiv1$ sets the units of length in the problem. We note that if $\lambda$ remains finite as $N\to\infty$, coherences are suppressed before the diffusive scale is reached. On the other hand, there exists a mesoscopic scaling regime, with $\lambda$ appropriately scaled at large $N$, in which quantum coherence may survive at large scale, and possibly coexists with interactions. 

At the single-replica level, the model reproduces the structure expected from MFT. Denoting by $\bar n(x;\tau)$ the averaged density, we find that, for $\lambda={\cal O}(1)$, $\bar n$ obeys a nonlinear diffusion equation with a density-dependent diffusion constant $D[\bar n]$, fixed explicitly by the microscopic interaction. The associated density fluctuations have the MFT form, with mobility $\sigma[\bar n]=D[\bar n]\,2\bar n(1-\bar n)$. On the contrary, as $\lambda \to 0$,  interactions become too weak to change the leading diffusive density sector, which reduces to the (Q)SSEP hydrodynamics~\cite{Derrida2007,Bernard2019}, $D[\bar n]\to D_0$. Thus, at the level of single-replica observables, weak interactions are invisible at diffusive scales.

The higher-replica sector is more sensitive. The interacting coherent regime of interest can be interpreted as a renormalization-group scaling statement around the QSSEP dynamics. Indeed, weak interactions act as a perturbation on the replicated QSSEP dynamics, whose coherence length is infinite: $\xi(\lambda=0)\approx \infty$. In this analogy, QSSEP plays the role of the critical fixed point, while $\lambda$ controls the departure away from criticality, similar to temperature variations $|T-T_c|$ in the Ising model~\cite{Cardy1996}. In a way analogous to field theory description of near-critical systems, we then tune the coupling $\lambda$ such that the coherence length $\xi(\lambda)$ remains finite as $N\to\infty$, resulting in the scaling $\lambda\sim \lambda_*$, with $\lambda_*=N^{-1+\alpha/2}$. As discussed below, for $\lambda\sim \lambda_*$, the interaction survives as a finite density-dependent mass term in the hydrodynamic-like equations for the coherence loops. Thus, coherence loop behavior detects interactions which are irrelevant for the density field dynamics.
If $\lambda\ll \lambda_*$, i.e. if $\lambda$ vanishes too rapidly with $N$, the perturbation becomes irrelevant also in the coherence sector and the QSSEP results are recovered.  For $\lambda \gg \lambda_*$, the system is instead incoherent at macroscopic scales. The scaling diagram of IQSEP summarized in Fig.~\ref{fig:illustration}(b) is the first main result of the Letter. Interactions strong enough to produce nonlinear diffusion also suppress coherence at diffusive scales. But, interactions too weak to affect the single-replica hydrodynamics may still control the behavior of higher-replica coherence loops. The resulting hydrodynamic-like equation for the coherence loops, see Eq.~\eqref{eq:g2-massive}, is the second main result.

\begin{figure}[t]
\centering
\includegraphics[width=\columnwidth]{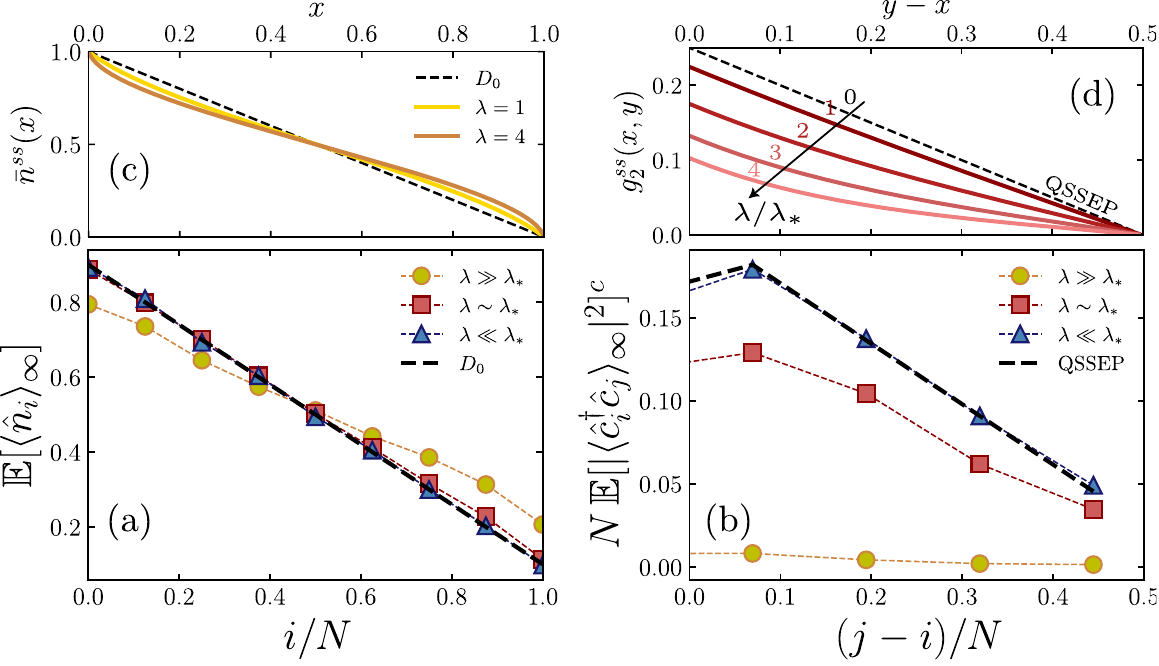}
\caption{Numerical checks of the IQSEP scaling picture. Panels (a)-(b) show exact-diagonalization data for the microscopic model \eqref{eq:dH-iqsep} with $N=8$ links, short-range dressing \eqref{eq:short-range-dressing}, and $D_0=1$. We plot the steady-state density (a) and coherence loop (b), as functions of $i/N$ with $j=4$ fixed in (b). The couplings are $\lambda=2$ (circ.), $\lambda=2/N$ (squares), and $\lambda=2/N^2$ (triang.), with $\lambda_*=1/N$. Boundary densities are $n_a=1$, $n_b=0$. Dashed lines show the finite-$N$ QSSEP result of Ref.~\cite{Bernard2019}. Data are averaged over $500$ trajectories, and evolved up to $t=2N^2$ with $dt\sim10^{-2}$. Panels (c)-(d) show the corresponding steady-state solutions of the large-scale IQSEP equations for the density (c)~(eq.~\eqref{eq:dens-eq}) and coherence loop (d)~(eq.~\eqref{eq:g2-massive}).}\label{fig:num}
\end{figure}

We have performed exact diagonalization numerics for the microscopic model illustrated in Fig.~\ref{fig:illustration}(a). The numerical data shown in Fig.~\ref{fig:num}(a)-(b) support the physical picture of Fig.~\ref{fig:illustration}(b). Moreover, despite the modest system size used ($N=8$), the results are in qualitative agreement with the large-scale predictions shown in Fig.~\ref{fig:num}(c)-(d) and discussed below. \\

\prlsection{Model and single-replica hydrodynamics}
We now introduce the model more precisely. The microscopic dynamics is generated on a one-dimensional chain of $N$ links by the stochastic Hamiltonian increment
\begin{equation}
d\hat H_t:=\sum_{k=0}^{N-1}\hat P_k \left(\hat\ell_k\,dW^k_t+\hat\ell_k^\dagger\,d\overline W^k_t\right),
\label{eq:dH-iqsep}
\end{equation}
 with $\hat\ell_k:=\hat c_{k+1}^\dagger\hat c_k$ where $\hat c^\dagger_j,\hat c_j$ are spinless Fermi operators. The stochastic increments $dW_t^k$, one per link, have zero mean and quadratic variation $dW_t^k d\overline W_t^l=\delta_{kl}dt$, in standard It\=o notation. For $\lambda=0$, i.e. in absence of interaction, the model reduces to the standard QSSEP of Refs.~\cite{Bauer2019,Bernard2019,Hruza2023}. The interaction is diagonal in the occupation basis. Hence it commutes with all densities,
$[\hat P_k,\hat n_j]=0$, and the stochastic dynamics preserves the local $U(1)$ symmetry. Without loss of generality, we choose a representative $\hat P_k$ which does not depend on the occupations $\hat n_k$ and $\hat n_{k+1}$ around the hopping link, namely such that $\{0, 1\}\not\in S_N$. With this convention, $[\hat P_k,\hat\ell_k]=[\hat P_k,\hat\ell_k^\dagger]=0$. The general form considered is therefore,
\begin{equation}
\Pt_k=\sum_{r= 1}^{|S_N|}\frac{b_r}{|S_N|^r}\sum_{\substack{Q\subset S_N;\\ |Q|=r}}\; \prod_{q\in Q}\hat n_{k+q},
    \label{eq:P-subext}
\end{equation}
where the coefficients $b_r$ depend only on the number $r=|Q|$ of occupations entering the monomials. The finite-range case $\alpha=0$ is obtained by keeping $S_N$ fixed and gives a finite polynomial in the neighbouring occupations. Imposing particle-hole symmetry, one may take, for instance,
\begin{equation}
\Pt_k\big\vert_{\alpha=0}=\hat n_{k-1}(1-\hat n_{k+2})+\hat n_{k+2}(1-\hat n_{k-1}) .
    \label{eq:short-range-dressing}
\end{equation}
Eq.~\eqref{eq:short-range-dressing} already captures the basic mechanism while keeping the results more compact, and will therefore serve as a reference example throughout the Letter. This choice also describes the effective slow dynamics of the noisy XXZ spin chain at strong coupling~\cite{Bauer2017}, recently investigated in Ref.~\cite{noisyXXZ-2026}.
 
 The evolution of the system density matrix, which is unitary (up to boundary terms) but stochastic, is obtained from $\hat\rho_{t+dt}=e^{-\I d\hat H_t}\hat\rho_t e^{\I d\hat H_t}$, and gives
\begin{equation}
d\hat\rho_t=-\I[d\hat H_t,\hat\rho_t]-\frac12[d\hat H_t,[d\hat H_t,\hat\rho_t]]+\text{bdry}.
    \label{eq:rho-dynamics}
\end{equation}
Its explicit form can be read in the \hyperlink{EM}{End Matter}. Boundary injection and extraction can be added as in QSSEP, see~\cite{Bernard2019,Hruza2023}. In the hydrodynamic limit they impose the boundary densities
$\bar n(0)=n_a$ and $\bar n(1)=n_b$, while the bulk equations derived below
are unchanged. This is the setting used in the numerical checks of Fig.~\ref{fig:num}. See the Supplemental Material (\hyperlink{SM}{SM}) for details.

Upon noise-averaging, only the second term, $-\frac 12[d\hat H_t,[d\hat H_t,\bullet]]$, of the unitary dynamics in \eqref{eq:rho-dynamics} survives. The latter is in the form of a dissipative Lindbladian, and projects $\bar\rho_t:=\E[\hat\rho_t]$ onto a diagonal subspace exponentially fast in time \cite{Bernard2018}: $\bar\rho_t\approx \sum_\mathbf{n} \Pi_t(\mathbf{n})|\mathbf{n}\rangle\langle\mathbf{n}|$ with $\Pi_t(\mathbf{n})$ a probability distribution on classical configurations $\mathbf{n}=\big(n_0,\dots,n_N\big)$ of the chain. The noise-averaged quantum dynamics becomes a classical Markov chain with master equation,
\be\label{eq:ME}
\frac{d}{dt}\Pi_t(\mathbf{n})=\sum_k\big(\gamma^+_k \Pi_t(\mathbf{n}_k^+)+ \gamma^-_k \Pi_t(\mathbf{n}_k^-)- (\gamma_k^++\gamma_k^-)\Pi_t(\mathbf{n}) \big),
\ee
with transition rates
\begin{align}\label{eq:trans-rates-ME}
&\gamma_k^+=\langle \mathbf{n}|\hat D_k|\mathbf{n}\rangle\,  n_k(1- n_{k+1}),\nn
&\gamma_k^-=\langle \mathbf{n}|\hat D_k|\mathbf{n}\rangle\,  n_{k+1}(1- n_{k}),
\end{align}
with $\hat D_k:=\hat P_k^2$, and configurations $\mathbf{n}_k^\pm$ obtained from $\mathbf{n}$ by moving a particle from site $k$ to the right/left neighbor site. Eqs.~\eqref{eq:ME}-\eqref{eq:trans-rates-ME} define a classical  interacting symmetric exclusion process, as discussed in literature, e.g. in Refs.~\cite{Funaki1991,Gonalves2009,Blondel2020,DaCunha2026}.

We write $\langle\!\langle\bullet\rangle\!\rangle_t:=\mathbb E[\mathrm{tr}(\hat\rho_t\,\bullet)]$. Since density observables
commute with the interaction dressing, their one-replica evolution is encoded in the classical interacting exclusion process. In particular, the above-mentioned local $U(1)$ symmetry ensures a local conservation law, $\partial_t\langle\!\langle \hat n_i\rangle\!\rangle_t+\langle\!\langle \hat J_i-\hat J_{i-1}\rangle\!\rangle_t=0$, with local current
\begin{equation}
\hat J_i:=-\hat D_i\,(\hat n_{i+1}-\hat n_i).
\label{eq:microscopic-current}
\end{equation}
As expected, the interaction only dresses the local jump rate in a way depending on the local density environment. Taking the diffusive scaling limit introduced above, we define
\begin{equation}
    \bar n(x,\tau):=\lim_{N\to\infty}\langle\!\langle \hat n_i\rangle\!\rangle_t\; \Big\vert_{\substack{x=i/N \\ \tau =t/N^2}} .
    \label{eq:nbar-def}
\end{equation}
Since Eq.~\eqref{eq:ME} codes for a classical diffusive Markov chain, the density cumulants are expected to fulfil to the standard MFT scaling:
$\langle\!\langle\hat O_n \rangle\!\rangle^c\sim N^{1-n}$ for $\hat O_n:= \hat n_{i_1} \dots \hat n_{i_n}$ and $i_q$ all distinct, see e.g. Ref.~\cite{Bertini2015}. Using MFT scaling at large $N$ yields to the closure of the otherwise infinite hierarchy of equations for the density correlation functions.   For instance, one finds 
\be\label{eq:mft-factorization}
\langle\!\langle\Pt_k \hat{O}_n\rangle\!\rangle=\langle\!\langle \Pt_k  \rangle\!\rangle \langle\!\langle\hat{O}_n\rangle\!\rangle+{\cal O}(N^{\alpha-1}),
\ee
for $i_1,\dots, i_n\not\in k+S_N$. Eq.~\eqref{eq:mft-factorization} follows by noticing that the cumulant $\Cum(\hat O_n,\hat O_m):=\avg{\hat O_n \ \hat O_m} -\avg{\hat O_n}\avg{\hat O_m}$ is at least ${\cal O}(1/N)$ if the supports of the two density operators $\hat O_n, \hat O_m$ do not overlap.

The current $\avg{\hat J_i}$ in \eqref{eq:microscopic-current} then closes at leading order in $1/N$, and yields the diffusion equation
\be\label{eq:dens-eq}
\partial_\tau \bar n=\partial_x \left(D[\bar n]\partial_x\bar n\right), 
\ee
with $\bar{n}(0)=n_a$, $\bar{n}(1)=n_b$ and diffusivity 
\be
D(x;\tau):=\lim_{N\to\infty}\avg{\hat D_i}_{t}\; \Big\vert_{\substack{x=i/N\\ \tau=t/N^2}}\equiv D[\bar n(x;\tau)]
\ee
given by
\be
D[\bar n]=\sum_{m\geq 0} B_m \bar n^m,\; B_m:=\!\!\!\!\sum_{\substack{Q,Q'\subset S_N \\ |Q\cup Q'|=m}}\frac{D_0\lambda^2 b_{|Q|} b_{|Q'|}}{|S_N|^{|Q|+|Q'|}},
\label{eq:Diff-const}
\ee
with $b_0= 1/\lambda$. The finite-range case gives a finite polynomial. From Eq.~\eqref{eq:short-range-dressing}, $D[\bar n]= D_0\left[1+2(2\lambda+\lambda^2)\bar n(1-\bar n)\right]$. Therefore the one-replica sector already separates two regimes. If $\lambda={\cal O}(1)$, the interaction survives at large scales and produces a nonlinear density hydrodynamics. If $\lambda\to 0$, then $D[\bar n]\to D_0$ and the leading density hydrodynamics reduces to that of (Q)SSEP. The numerical steady-state solution of \eqref{eq:dens-eq} for the interaction~\eqref{eq:short-range-dressing} is shown in Fig.~\ref{fig:num}(c). Details of the derivations are given in the \hyperlink{SM}{SM}.

\indent
A similar calculation fixes the mobility. We introduce the scaling function of the density-density cumulant, $C_2(x,y;\tau):=\lim_{N\to\infty}N\,\langle\!\langle\hat n_i\hat n_j\rangle\!\rangle_t^c$, with $x=i/N$, $y=j/N$, and $\tau=t/N^2$. In this scaling limit, one obtains, see \hyperlink{SM}{SM},
\begin{equation}
    \left(\de_\tau-\Delta_x^D-\Delta_y^D\right)C_2(x,y;\tau)=\de_x\de_y\left[\sigma(\bar n(x,\tau))\delta(x-y)\right],
    \label{eq:C-hydro}
\end{equation}
with $\Delta_x^D\bullet:=\partial_x^2\left(D[\bar n(x,\tau)]\,\bullet\right)$, and
\begin{equation}
    \sigma(\bar n)=D[\bar n]\,\sigma_0(\bar n),
    \label{eq:mobility}
\end{equation}
with $\sigma_0(\bar n)=2\bar n(1-\bar n)$, and boundary conditions $C_2(0, y;\tau) = 0$, $C_2(1, y;\tau)=0$, and similarly for $y$ near the boundaries.

The interaction changes the transport coefficient but not the local equilibrium thermodynamics. Eq.~\eqref{eq:mobility} is consistent with the structure of the
noise-averaged process. At equilibrium, the latter satisfies detailed balance with respect to the uniform measure at fixed particle number. Hence the local equilibrium free-energy density is not modified by the interactions, and remains $F(\bar n)=\bar n\log \bar n+(1-\bar n)\log(1-\bar n)$. The mobility obeys the Einstein relation $\sigma(\bar n)=2D[\bar n]/F''(\bar n)=D[\bar n]\,2\bar n(1-\bar n)$, in agreement with \eqref{eq:mobility}; see \hyperlink{SM}{SM} for further details. 

Scaling functions of higher density correlations, $\avg{\hat n_{i_1}\dots\hat n_{i_k}}^c_t$, can be computed analogously, or read directly from the associated MFT description~\cite{Bertini2015}. This completes the single-replica part of the theory.
\\

\prlsection{Multi-replica coherences.}
The next question is whether multi-replica coherences, which are not captured by the single-replica MFT,  remain visible on the same diffusive scale. By the local $U(1)$ symmetry, charged one-replica
observables vanish exponentially fast after noise averaging; in particular $\langle\!\langle \hat c_i^\dagger \hat c_j\rangle\!\rangle_{t\gg1}=0$ for $i\neq j$. Information on quantum coherences is therefore carried by neutral multi-replica observables. The first of such object is the two-replica coherence loop,
\begin{equation}
    g_2(x,y;\tau):=\lim_{N\to\infty}N\,\mathbb E\left[\langle \hat X_{ij}\rangle_t
    \ \langle \hat X_{ji}\rangle_t\right]^c,
    \label{eq:g2-def}
\end{equation}
with $\hat X_{ij}:=\hat c_i^\dagger\hat c_j$, $x=i/N$, $y=j/N$, and $\langle \bullet\rangle_t:=\tr(\hat\rho_t \ \bullet)$. This object gives a simple measure of the coherences which are invisible
at the single-replica level. Its equation of motion has two distinct contributions. The terms already
present in QSSEP, which generate diffusion and a contact source, see~\cite{Hruza2023}. The additional terms generated by the interaction, which act as a dephasing mass for the loop. To leading order in $1/N$, these additional terms have the structure (see the \hyperlink{SM}{SM}),
\begin{align}
&a^{(1)}_{i,t}:=2N^2\big(\avg{\hat D_i}_t-\avg{\hat P_i}_t^2\big)={\cal O}(\lambda^2/\lambda_*^2);\nn
&a^{(2)}_{i,t}:=N^2\sum_k\avg{( \de_{\hat n_i} \hat P_k)^2(\hat n_k-\hat n_{k+1})^2}_t={\cal O}(\lambda^2/\lambda_*^2).
    \label{eq:mass-counting}
\end{align}
For $N\to\infty$, these terms are finite only if $\lambda$ is scaled appropriately with $N$, i.e., precisely when $\lambda\sim\lambda_*$. This defines the {\it mesoscopic scaling regime}. Although irrelevant in the single-replica hydrodynamics for $\lambda\to 0$, in the scaling regime the interaction remains visible in the two- or higher-replica sectors, with a finite coherence length in the continuum. In the mesoscopic scaling regime, the two-replica coherence loop satisfies a massive
hydrodynamic-like equation of the form 
\be
\big(\de_\tau-D_0\sum_{r=x,y}[\de_r^2-m^2(r;\tau)]\big)g_2(x,y;\tau)={\cal S}^{\rm qssep}_2(x,y;\tau),
\label{eq:g2-massive}
\ee
where ${\cal S}^{\rm qssep}_2=2D_0\delta(x-y)(\partial_x \bar n)^2$ is the contact source already present in
QSSEP~\cite{Hruza2023}, and the density-dependent square mass is
\be\label{eq:mass}
m^2(x;\tau):=D_0^{-1}\lim_{N\to\infty}\big(  a^{(1)}_{i,t}+a^{(2)}_{i,t}\big)\Big\vert_{\substack{x=i/N \\ \tau=t/N^2}}=\sum_{m\geq 1} \!\! A_m \bar n(x,\tau)^m.
\ee
The explicit coefficients $A_m$ in terms of the interaction parameters $b_r$ are given in the \hyperlink{SM}{SM}. For the example of Eq.~\eqref{eq:short-range-dressing}, this
reduces to $m^2(\bar n)= 2c_\lambda\sigma_0(\bar n)\big[2-\sigma_0(\bar n)\big]$, with $c_\lambda=\lambda^2/\lambda^2_*$. The boundary conditions are $g_2(0,y;\tau)=0$ , $g_2(1,y;\tau)=0$ and similarly for $y$. 
The derivation of Eq.~\eqref{eq:g2-massive} is given in the \hyperlink{SM}{SM}.  Its steady-state numerical solution for the interaction~\eqref{eq:short-range-dressing} is shown in Fig.~\ref{fig:num}(d). 
The mass term in Eq.~\eqref{eq:g2-massive} sets the local coherence length $\xi\sim 1/m(x;\tau)$, which is thus density dependent. At length scales smaller than $1/m(x;\tau)$, the system behavior is coherent and governed by the QSSEP, while decoherence takes place at scales bigger than $1/m(x;\tau)$ and the system there behaves classically. The continuous scaling theory describes this crossover.

The two-replica calculation extends naturally to higher-replica coherence loops. The same power-counting in Eq.~\eqref{eq:mass-counting}  shows that the interaction scaling $\lambda\sim\lambda_*$ is independent of the replica number. Higher coherence loops satisfy equations similar to Eq.~\eqref{eq:g2-massive}. One finds the large-scale equation of Ref.~\cite{Hruza2023} with the additional square mass \eqref{eq:mass} on each point forming the loop.  We refer to  the \hyperlink{SM}{SM} for details on this generalization.\\

\prlsection{Deviation from Gaussianity.}
We now discuss how much the IQSEP dynamics deviates from the realization-wise Gaussian behavior of QSSEP due to the presence of interactions. We show that the interaction-induced mass \eqref{eq:mass} and the coherence loops control this departure. For two replicas, we quantify this deviation by introducing the quantity $\Gamma_2(x,y;\tau):=C_2(x,y;\tau)+g_2(x,y;\tau)-\bar n(x;\tau)(1-\bar n(x;\tau))\delta(x-y)$, which would vanish identically for a Gaussian state. Eqs.~(\ref{eq:C-hydro},\ref{eq:g2-massive}) then imply (see the \hyperlink{SM}{SM}),
\begin{equation}
 \big(\de_\tau-D_0(\de_x^2+\de_y^2)\big)\Gamma_2(x,y)=-D_0\big(m^2(x)+m^2(y)\big)g_2(x,y).
    \label{eq:Gamma-hydro}
\end{equation}
The non-Gaussian correction is thus slaved to the coherence loop and to the same mass that controls its dephasing. In particular, it disappears in the QSSEP limit where $m^2\to 0$. Similar statements apply to higher-point correlation functions, by introducing appropriate generalizations of $\Gamma_2$ coding for the irreducible components of correlation functions as in field theory.
\\

\prlsection{Computational strategy.}
We finally briefly state the main strategy and assumption underlying the derivation of the results we discussed. The factorization property in Eq.~\eqref{eq:mft-factorization} follows from MFT and, noticeably, closes the hierarchy of equations of motion in the single-replica sector. MFT is a large deviation theory, as usual in statistical physics, so that its scaling relations echo that large density fluctuations are rare and irreducible density correlations are subleading in the system size. However, MFT scaling is not sufficient to close the hierarchy of equations in the higher replica sectors. For instance, the two-replica dynamics generates mixed objects in which density dressings multiply charged operators, such as $\E[\langle \hat n_k\hat X_{ij}\rangle_t\langle \hat X_{ji}\rangle_t]$. Therefore, following MFT logic, which is expected to also hold in the higher replica sectors, we assume that density fluctuations are subleading in $1/N$ so that, away from contact points, density dressings factorize locally against the coherence loops at leading order. We checked this assumption self-consistently from the equations of motion. In its simplest form it states that 
\be\label{eq:factorization}
\E[ \langle \hat n_k\hat X_{ij}\rangle_t \langle \hat X_{ji}\rangle_t] = \langle\!\langle \hat n_k\rangle\!\rangle_t\, \E[\langle \hat X_{ij}\rangle_t\langle \hat X_{ji}\rangle_t]+{\cal O}(N^{-2}),
\ee
for $k\neq i,j$, and with analogous formulas for higher density polynomials inserted on the two replicas. Eq.~\eqref{eq:factorization} might be seen as the natural counterpart of the density factorization used in the single-replica MFT sector; its consistency is discussed in the \hyperlink{EM}{End Matter}.\\

\prlsection{Conclusion.}
The IQSEP thus provides a simple, versatile, and analytically tractable family of models coding for interacting, fluctuating, quantum diffusive transport. By incorporating a tunable coherence length, the mesoscopic scaling regime we introduced offers an effective description of the crossover from coherent diffusive transports and their fluctuations at small scales to incoherent ones at large scales. While this scaling theory shares conceptual similarities with effective field theory, it transcends the limitations of standard fluctuating hydrodynamics, which only captures the one-replica sector.   
These findings pave the way for a quantum extension of macroscopic fluctuation theory to the mesoscopic coherent sector, namely a quantum mesoscopic fluctuation theory (QMFT) for interacting diffusive quantum systems~\cite{Bernard2021}.

The IQSEP model is also relevant beyond the scope of QMFT, as an effective description of large-scale transport in noisy interacting spin chains. In particular, the parallel work~\cite{noisyXXZ-2026} investigates the XXZ spin chain with dephasing noise and shows that, in the strong-coupling regime, its slow dynamics is described by an IQSEP of the form considered in Eq.~\eqref{eq:short-range-dressing}. The resulting predictions are then compared with tensor-network simulations.

A natural next step would be to extend the continuum formulation of the QSSEP~\cite{Bernard2026-qssep-cont} to the IQSEP, thereby establishing a direct continuum framework for the QMFT. Such a theory is expected to apply broadly to noisy fermionic systems with diffusive transport and would provide a route to large-deviation principles for rare quantum-coherent fluctuations, beyond the cumulant expansion considered in this work.
Another promising direction involves generalizing this class of stochastic quantum diffusion to model systems with diverse equilibrium free energies.  We intend to explore these questions in the not-too-distant future. 


\medskip
\prlsection{Acknowledgements.} We acknowledge Tony Jin, Adam Nahum, Ewan McCulloch, and Fabian Essler for useful discussions and feedback on the project. SS and DB also acknowledge Alexios Christopoulos, Jo\~ao Costa, Jacopo De Nardis,
Tony Jin, and Zala Lenar\v{c}i\v{c} for recent collaboration on related topics. SS is supported by the MSCA Grant No.~101103348 (GENESYS). FH has received support under the Major Research Program of PSL Research University ``Statistical Physics and Mathematics'' launched by PSL Research University and implemented by ANR with the references ANR-10-IDEX-0001. DB is partly supported by the CNRS, the ENS and the Simons Foundation via the Simons Collaboration on Probabilistic Paths to QFT. Views and opinions expressed are those of the authors only and do not necessarily reflect those of the European Union or the European Research Council Executive Agency. Neither the European Union nor the granting authority can be held responsible for them.

\bigskip\bigskip

\begin{center}
\hypertarget{EM}{\Large \textbf{End Matter}}
\end{center}
\newcounter{emsection}
\newcommand{\emsection}[1]{%
  \refstepcounter{emsection}%
  \section*{\Roman{emsection}. #1}%
}
\emsection{IQSEP Stochastic evolution}
The starting point of our derivations is the stochastic differential equation satisfied by a generic operator $\hat O$ in the Heisenberg picture. Using Eq.~\eqref{eq:dH-iqsep} in the adjoint of Eq.~\eqref{eq:rho-dynamics}, one obtains
\be
d\hat O= \sum_{k=0}^{N-1}\big( \I d{\cal H}_k(\hat O) + dt \ {\cal L}_k(\hat O)\big),
\ee
with stochastic increment $d{\cal H}_k(\bullet):=[\hat P_k d\hat{h}_k, \bullet]$, and $d\hat{h}_k:= \hat\ell_k dW_t^k +\hat\ell^\dagger_k d\overline{W}_t^k$ is the standard QSSEP martingale increment, see e.g. Refs.~\cite{Bauer2017,Bernard2019,Hruza2023}. The drift part ${\cal L}_k$ is deterministic in It\=o convention. 
Its local action can be written as
\be
{\cal L}_k(\bullet):= \hat P_k {\cal L}_k^\text{qssep}(\bullet )\hat P_k - {\cal D}_k(\bullet)
\ee
where, for convenience, we identified the QSSEP drift ${\cal L}_k^\text{qssep}(\bullet):=\hat\ell_k\bullet \hat \ell_k^{\dagger} + \hat\ell_k^\dagger \bullet \hat \ell_k -\frac12\big\{\hat \Delta_k ,\bullet \big\}$ and the additional drift correction
\be\label{eq:drift-correction}
{\cal D}_k(\bullet):=\frac{1}{2} [\hat P_k\hat\Delta_k,[\hat P_k,\bullet]].
\ee
Here, $\hat\Delta_k:=\big(\hat n_{k+1}-\hat n_k\big)^2$. Note that, for density operators $\hat O_n=\hat n_{i_1}\dots \hat n_{i_n}$, Eq.~\eqref{eq:drift-correction} vanishes and the noise-averaged dynamics reads simply,
\be\label{eq:single-rep-hydro}
\frac{d}{dt}\avg{\hat O_n}_t =\sum_{k=0}^{N-1} \avg{ \hat P_k {\cal L}_k^\text{qssep}(\hat O_n )\hat P_k }_t,
\ee
from which the single-replica MFT results can be derived using the MFT-factorization of Eq.~\eqref{eq:mft-factorization}; see \hyperlink{SM}{SM} for details. For the multi-replica sector, e.g. for the two-replica coherence loop, one has
\begin{align}\label{eq:2-replica}
\frac{d}{dt}\E[\langle\hat X_{ij}\rangle\langle \hat X_{ji}\rangle]&=\sum_k\E[\langle\hat X_{ij}\rangle\langle\big(\hat P_k{\cal L}^\text{qssep}_k(\hat X_{ji}) \hat P_k-{\cal D}_k(\hat{X}_{ji}\big)\rangle]\nn
&+(i\leftrightarrow j)-\frac{1}{dt}\sum_k \E[\langle d{\cal H}_k(\hat X_{ij})\rangle\langle d{\cal H}_k(\hat X_{ji})\rangle],
\end{align}
with the last term being the standard It\=o contraction arising from the stochastic dynamics on the two replicas. As shown in the \hyperlink{SM}{SM}, the It\=o correction and the non-vanishing drift correction ${\cal D}_k(\hat X_{ij})\neq 0$ entering \eqref{eq:2-replica} generate the contributions in \eqref{eq:mass-counting}, and ultimately the dephasing mass in Eq.~\eqref{eq:mass}.

\emsection{Self-consistency of the higher-replica factorization}\label{sec:factorization-hyp}
We now discuss the multi-replica factorization ansatz used in deriving our results. Our strategy is to bootstrap this ansatz directly from the equations of motion, while leaving a rigorous proof to future work.

To this end, we consider the two-replica coherence loop $\G_{ij}(t):=\E[\langle \hat X_{ij}\rangle_t\langle \hat X_{ji}\rangle_t]$, and 
\be
\G^{(A,B)}_{ij}(t):=\E[\langle \hat A \hat X_{ij}\rangle_t\langle \hat B \hat X_{ji}\rangle_t]
\ee
with $\hat A\equiv A(\hat n)$, $\hat B\equiv B(\hat n)$ some density polynomials. We then introduce the factorization error
\be\label{eq:Y-def}
Y^{(A,B)}_{\G}(t):=\G^{(A,B)}(t)-\avg{\hat A}_t\avg{\hat B}_t \G(t)
\ee
with scaling ansatz $Y^{(A,B)}_{\G}={\cal O}(N^{-\gamma})$, and $\gamma>\nu_\G+\nu_A+\nu_B$ to be determined self-consistently. Here, $\nu_\G=1$ is the scaling exponent of the two-replica loop, and
\be
\avg{\hat A}={\cal O}(N^{-\nu_A}); \quad \avg{\hat B}={\cal O}(N^{-\nu_B}).
\ee
For simplicity, we first take $\hat A$ and $\hat B$ whose supports are far apart, and separated from $\{i,i-1,j,j-1\}\cup (i-S_N)\cup(j-S_N)$. Such assumption is then relaxed by treating the resulting contact terms separately, see discussion below and \hyperlink{SM}{SM}.

 The time variation of $Y^{(A,B)}_\G$ is given by
\begin{align}\label{eq:dY}
dY^{(A,B)}_{\G}=&\Big(\E[ _A\langle d\hat X\rangle\  _B\langle \hat X^\dagger\rangle +  _A\langle \hat X\rangle\  _B\langle d\hat X^\dagger\rangle + _A\langle d \hat X\rangle\  _B\langle d \hat X^\dagger\rangle]\nn
&- \bar A\bar B( d\G)\Big) +\Big(\E[\langle d\hat A\rangle_X \ \langle \hat B\rangle_{X^\dagger}] - (d\bar{A})\bar B \G\Big)\nn
&+\Big( \E[\langle \hat A\rangle_X \ \langle d\hat B\rangle_{X^\dagger}] -\bar A (d\bar{B}) \G  \Big)+ R,
\end{align}
with shorthand $\avg{\hat A}\equiv\bar A$, same for $\bar B$, and with definition of the decorated quantum expectations,
\be
_A\langle \bullet\rangle:= \langle \hat{A} \bullet\rangle; \qquad \langle \bullet\rangle_X:= \langle  \bullet \hat{X}\rangle.
\ee
The terms in parentheses of \eqref{eq:dY} have the same structure as Eq.~\eqref{eq:Y-def}. Assuming the same factorization scaling for these terms,
\begin{align}
dY^{(A,B)}_{\G}=& Y_{d\G}^{(A,B)}+Y_{\G}^{(dA,B)}+ Y_{\G}^{(A,dB)} + R\nn
=& R + {\cal O}(N^{-\gamma-2}dt).
\end{align}
The remainder is 
\begin{align}\label{eq:R}
 R:=&\E[ \langle (d\hat A)\hat X\rangle \ \langle (d\hat B)\hat X^\dagger\rangle+\langle (d\hat A)(d\hat X)\rangle \langle \hat B \hat X^\dagger\rangle\nn
&+ \langle \hat A\hat X\rangle \langle (d\hat B)(d\hat X^\dagger)\rangle + \langle (d\hat A)\hat X\rangle \langle \hat B (d\hat X^\dagger)\rangle\nn
&+ \langle \hat A(d\hat X)\rangle \langle (d\hat B)\hat X^\dagger\rangle].
\end{align}
The quadratic variations entering \eqref{eq:R} are given by the martingales $d\hat A=\I\sum_k [\hat P_k d\hat h_k, \hat A]+{\cal O}(dt)$, and similarly for the others. These terms have support on links $k\ :\ \{k,k+1\}\cap \text{supp}(\hat A)\neq \emptyset$, same for $\hat B$, while $\I\sum_k [\hat P_k d\hat h_k, \hat X_{ij}]$ is supported in the region $\{i,i-1,j,j-1\}\cup (i-S_N)\cup(j-S_N)$. For well-separated density polynomials, the martingale supports do not overlap; therefore, the It\=o contractions all vanish, and $R=0$. Hence, away from contact configurations,  the error $Y_\G^{(A,B)}$ is dynamically stable.

The value of $\gamma$ is specified by the contacts. One finds that the dominant contact contribution scales as
\be\label{eq:Y-contact}
dY^{(A,B)}_\G\Big\vert_\text{contact}={\cal O}(N^{-3-(\nu_A+\nu_B+\nu_\G)} dt).
\ee
This scaling follows from the delta condition on contact configurations (contributing as $1/N$ in the hydrodynamic variables), together with the emergence of a double-gradient structure. Details of the derivation can be found in \hyperlink{SM}{SM}. Eq.~\eqref{eq:Y-contact} yields,
\be
\gamma \geq 1+ \nu_A+\nu_B+\nu_\G.
\ee
For the example in Eq.~\eqref{eq:factorization}, $\gamma\geq 2$.

Generalization to coherence loops made of $n>2$ points proceeds analogously, upon replacing $\nu_\G$ with $n-1$.

\bibliography{biblio_v2}

\begin{thebibliography}{63}%
\makeatletter
\providecommand \@ifxundefined [1]{%
 \@ifx{#1\undefined}
}%
\providecommand \@ifnum [1]{%
 \ifnum #1\expandafter \@firstoftwo
 \else \expandafter \@secondoftwo
 \fi
}%
\providecommand \@ifx [1]{%
 \ifx #1\expandafter \@firstoftwo
 \else \expandafter \@secondoftwo
 \fi
}%
\providecommand \natexlab [1]{#1}%
\providecommand \enquote  [1]{``#1''}%
\providecommand \bibnamefont  [1]{#1}%
\providecommand \bibfnamefont [1]{#1}%
\providecommand \citenamefont [1]{#1}%
\providecommand \href@noop [0]{\@secondoftwo}%
\providecommand \href [0]{\begingroup \@sanitize@url \@href}%
\providecommand \@href[1]{\@@startlink{#1}\@@href}%
\providecommand \@@href[1]{\endgroup#1\@@endlink}%
\providecommand \@sanitize@url [0]{\catcode `\\12\catcode `\$12\catcode
  `\&12\catcode `\#12\catcode `\^12\catcode `\_12\catcode `\%12\relax}%
\providecommand \@@startlink[1]{}%
\providecommand \@@endlink[0]{}%
\providecommand \url  [0]{\begingroup\@sanitize@url \@url }%
\providecommand \@url [1]{\endgroup\@href {#1}{\urlprefix }}%
\providecommand \urlprefix  [0]{URL }%
\providecommand \Eprint [0]{\href }%
\providecommand \doibase [0]{https://doi.org/}%
\providecommand \selectlanguage [0]{\@gobble}%
\providecommand \bibinfo  [0]{\@secondoftwo}%
\providecommand \bibfield  [0]{\@secondoftwo}%
\providecommand \translation [1]{[#1]}%
\providecommand \BibitemOpen [0]{}%
\providecommand \bibitemStop [0]{}%
\providecommand \bibitemNoStop [0]{.\EOS\space}%
\providecommand \EOS [0]{\spacefactor3000\relax}%
\providecommand \BibitemShut  [1]{\csname bibitem#1\endcsname}%
\let\auto@bib@innerbib\@empty
\bibitem [{\citenamefont {Liggett}(1985)}]{Liggett1985}%
  \BibitemOpen
  \bibfield  {author} {\bibinfo {author} {\bibfnamefont {T.~M.}\ \bibnamefont
  {Liggett}},\ }\href {https://doi.org/10.1007/978-1-4613-8542-4} {\emph
  {\bibinfo {title} {Interacting Particle Systems}}}\ (\bibinfo  {publisher}
  {Springer New York},\ \bibinfo {year} {1985})\BibitemShut {NoStop}%
\bibitem [{\citenamefont {Spohn}(1991)}]{Spohn1991}%
  \BibitemOpen
  \bibfield  {author} {\bibinfo {author} {\bibfnamefont {H.}~\bibnamefont
  {Spohn}},\ }\href {https://doi.org/10.1007/978-3-642-84371-6} {\emph
  {\bibinfo {title} {Large Scale Dynamics of Interacting Particles}}}\
  (\bibinfo  {publisher} {Springer Berlin Heidelberg},\ \bibinfo {year}
  {1991})\BibitemShut {NoStop}%
\bibitem [{\citenamefont {Kipnis}\ and\ \citenamefont
  {Landim}(1999)}]{Kipnis1999}%
  \BibitemOpen
  \bibfield  {author} {\bibinfo {author} {\bibfnamefont {C.}~\bibnamefont
  {Kipnis}}\ and\ \bibinfo {author} {\bibfnamefont {C.}~\bibnamefont
  {Landim}},\ }\href {https://doi.org/10.1007/978-3-662-03752-2} {\emph
  {\bibinfo {title} {Scaling Limits of Interacting Particle Systems}}},\ Vol.\
  \bibinfo {volume} {320}\ (\bibinfo  {publisher} {Springer Berlin
  Heidelberg},\ \bibinfo {year} {1999})\BibitemShut {NoStop}%
\bibitem [{\citenamefont {Kipnis}\ \emph {et~al.}(1989)\citenamefont {Kipnis},
  \citenamefont {Olla},\ and\ \citenamefont {Varadhan}}]{Kipnis1989}%
  \BibitemOpen
  \bibfield  {author} {\bibinfo {author} {\bibfnamefont {C.}~\bibnamefont
  {Kipnis}}, \bibinfo {author} {\bibfnamefont {S.}~\bibnamefont {Olla}},\ and\
  \bibinfo {author} {\bibfnamefont {S.~R.~S.}\ \bibnamefont {Varadhan}},\
  }\bibfield  {title} {\bibinfo {title} {Hydrodynamics and large deviation for
  simple exclusion processes},\ }\href {https://doi.org/10.1002/cpa.3160420202}
  {\bibfield  {journal} {\bibinfo  {journal} {Commun. Pure Appl. Math.}\
  }\textbf {\bibinfo {volume} {42}},\ \bibinfo {pages} {115–137} (\bibinfo
  {year} {1989})}\BibitemShut {NoStop}%
\bibitem [{\citenamefont {Bertini}\ \emph {et~al.}(2001)\citenamefont
  {Bertini}, \citenamefont {De~Sole}, \citenamefont {Gabrielli}, \citenamefont
  {Jona-Lasinio},\ and\ \citenamefont {Landim}}]{bertini2001fluctuations}%
  \BibitemOpen
  \bibfield  {author} {\bibinfo {author} {\bibfnamefont {L.}~\bibnamefont
  {Bertini}}, \bibinfo {author} {\bibfnamefont {A.}~\bibnamefont {De~Sole}},
  \bibinfo {author} {\bibfnamefont {D.}~\bibnamefont {Gabrielli}}, \bibinfo
  {author} {\bibfnamefont {G.}~\bibnamefont {Jona-Lasinio}},\ and\ \bibinfo
  {author} {\bibfnamefont {C.}~\bibnamefont {Landim}},\ }\bibfield  {title}
  {\bibinfo {title} {Fluctuations in stationary nonequilibrium states of
  irreversible processes},\ }\href
  {https://doi.org/https://doi.org/10.1103/PhysRevLett.87.040601} {\bibfield
  {journal} {\bibinfo  {journal} {Phys. Rev. Lett.}\ }\textbf {\bibinfo
  {volume} {87}},\ \bibinfo {pages} {040601} (\bibinfo {year}
  {2001})}\BibitemShut {NoStop}%
\bibitem [{\citenamefont {Bertini}\ \emph {et~al.}(2002)\citenamefont
  {Bertini}, \citenamefont {De~Sole}, \citenamefont {Gabrielli}, \citenamefont
  {Jona-Lasinio},\ and\ \citenamefont {Landim}}]{bertini2002macroscopic}%
  \BibitemOpen
  \bibfield  {author} {\bibinfo {author} {\bibfnamefont {L.}~\bibnamefont
  {Bertini}}, \bibinfo {author} {\bibfnamefont {A.}~\bibnamefont {De~Sole}},
  \bibinfo {author} {\bibfnamefont {D.}~\bibnamefont {Gabrielli}}, \bibinfo
  {author} {\bibfnamefont {G.}~\bibnamefont {Jona-Lasinio}},\ and\ \bibinfo
  {author} {\bibfnamefont {C.}~\bibnamefont {Landim}},\ }\bibfield  {title}
  {\bibinfo {title} {Macroscopic fluctuation theory for stationary
  non-equilibrium states},\ }\href {https://doi.org/10.1023/a:1014525911391}
  {\bibfield  {journal} {\bibinfo  {journal} {J. Stat. Phys.}\ }\textbf
  {\bibinfo {volume} {107}},\ \bibinfo {pages} {635–675} (\bibinfo {year}
  {2002})}\BibitemShut {NoStop}%
\bibitem [{\citenamefont {Bertini}\ \emph {et~al.}(2005)\citenamefont
  {Bertini}, \citenamefont {De~Sole}, \citenamefont {Gabrielli}, \citenamefont
  {Jona-Lasinio},\ and\ \citenamefont {Landim}}]{Bertini2005}%
  \BibitemOpen
  \bibfield  {author} {\bibinfo {author} {\bibfnamefont {L.}~\bibnamefont
  {Bertini}}, \bibinfo {author} {\bibfnamefont {A.}~\bibnamefont {De~Sole}},
  \bibinfo {author} {\bibfnamefont {D.}~\bibnamefont {Gabrielli}}, \bibinfo
  {author} {\bibfnamefont {G.}~\bibnamefont {Jona-Lasinio}},\ and\ \bibinfo
  {author} {\bibfnamefont {C.}~\bibnamefont {Landim}},\ }\bibfield  {title}
  {\bibinfo {title} {Current fluctuations in stochastic lattice gases},\ }\href
  {https://doi.org/10.1103/physrevlett.94.030601} {\bibfield  {journal}
  {\bibinfo  {journal} {Phys. Rev. Lett.}\ }\textbf {\bibinfo {volume} {94}},\
  \bibinfo {pages} {030601} (\bibinfo {year} {2005})}\BibitemShut {NoStop}%
\bibitem [{\citenamefont {Bertini}\ \emph {et~al.}(2015)\citenamefont
  {Bertini}, \citenamefont {De~Sole}, \citenamefont {Gabrielli}, \citenamefont
  {Jona-Lasinio},\ and\ \citenamefont {Landim}}]{Bertini2015}%
  \BibitemOpen
  \bibfield  {author} {\bibinfo {author} {\bibfnamefont {L.}~\bibnamefont
  {Bertini}}, \bibinfo {author} {\bibfnamefont {A.}~\bibnamefont {De~Sole}},
  \bibinfo {author} {\bibfnamefont {D.}~\bibnamefont {Gabrielli}}, \bibinfo
  {author} {\bibfnamefont {G.}~\bibnamefont {Jona-Lasinio}},\ and\ \bibinfo
  {author} {\bibfnamefont {C.}~\bibnamefont {Landim}},\ }\bibfield  {title}
  {\bibinfo {title} {Macroscopic fluctuation theory},\ }\href
  {https://doi.org/10.1103/revmodphys.87.593} {\bibfield  {journal} {\bibinfo
  {journal} {Rev. Mod. Phys.}\ }\textbf {\bibinfo {volume} {87}},\ \bibinfo
  {pages} {593–636} (\bibinfo {year} {2015})}\BibitemShut {NoStop}%
\bibitem [{\citenamefont {Derrida}\ \emph {et~al.}(2001)\citenamefont
  {Derrida}, \citenamefont {Lebowitz},\ and\ \citenamefont
  {Speer}}]{Derrida2001}%
  \BibitemOpen
  \bibfield  {author} {\bibinfo {author} {\bibfnamefont {B.}~\bibnamefont
  {Derrida}}, \bibinfo {author} {\bibfnamefont {J.~L.}\ \bibnamefont
  {Lebowitz}},\ and\ \bibinfo {author} {\bibfnamefont {E.~R.}\ \bibnamefont
  {Speer}},\ }\bibfield  {title} {\bibinfo {title} {Free energy functional for
  nonequilibrium systems: An exactly solvable case},\ }\href
  {https://doi.org/10.1103/PhysRevLett.87.150601} {\bibfield  {journal}
  {\bibinfo  {journal} {Phys. Rev. Lett.}\ }\textbf {\bibinfo {volume} {87}},\
  \bibinfo {pages} {150601} (\bibinfo {year} {2001})}\BibitemShut {NoStop}%
\bibitem [{\citenamefont {Derrida}(2007)}]{Derrida2007}%
  \BibitemOpen
  \bibfield  {author} {\bibinfo {author} {\bibfnamefont {B.}~\bibnamefont
  {Derrida}},\ }\bibfield  {title} {\bibinfo {title} {Non-equilibrium steady
  states: fluctuations and large deviations of the density and of the
  current},\ }\href {https://doi.org/10.1088/1742-5468/2007/07/P07023}
  {\bibinfo  {journal} {J. Stat. Mech.: Theory Exp.}\ ,\ \bibinfo {pages}
  {7023}}\BibitemShut {NoStop}%
\bibitem [{\citenamefont {Derrida}(2011)}]{Derrida2011}%
  \BibitemOpen
\bibfield  {journal} {  }\bibfield  {author} {\bibinfo {author} {\bibfnamefont
  {B.}~\bibnamefont {Derrida}},\ }\bibfield  {title} {\bibinfo {title}
  {Microscopic versus macroscopic approaches to non-equilibrium systems},\
  }\href {https://doi.org/10.1088/1742-5468/2011/01/p01030} {\bibfield
  {journal} {\bibinfo  {journal} {J. Stat. Mech.: Theory Exp.}\ }\textbf
  {\bibinfo {volume} {2011}}\bibinfo  {number} { (01)},\ \bibinfo {pages}
  {P01030}}\BibitemShut {NoStop}%
\bibitem [{\citenamefont {Mallick}(2015)}]{Mallick2015}%
  \BibitemOpen
\bibfield  {number} {  }\bibfield  {author} {\bibinfo {author} {\bibfnamefont
  {K.}~\bibnamefont {Mallick}},\ }\bibfield  {title} {\bibinfo {title} {The
  exclusion process: A paradigm for non-equilibrium behaviour},\ }\href
  {https://doi.org/10.1016/j.physa.2014.07.046} {\bibfield  {journal} {\bibinfo
   {journal} {Physica A}\ }\textbf {\bibinfo {volume} {418}},\ \bibinfo {pages}
  {17} (\bibinfo {year} {2015})}\BibitemShut {NoStop}%
\bibitem [{\citenamefont {Bernard}(2021)}]{Bernard2021}%
  \BibitemOpen
  \bibfield  {author} {\bibinfo {author} {\bibfnamefont {D.}~\bibnamefont
  {Bernard}},\ }\bibfield  {title} {\bibinfo {title} {Can the macroscopic
  fluctuation theory be quantized?},\ }\href
  {https://doi.org/10.1088/1751-8121/AC2597} {\bibfield  {journal} {\bibinfo
  {journal} {J. Phys. A: Math. Theor.}\ }\textbf {\bibinfo {volume} {54}},\
  \bibinfo {pages} {433001} (\bibinfo {year} {2021})}\BibitemShut {NoStop}%
\bibitem [{\citenamefont {Nahum}\ \emph {et~al.}(2018)\citenamefont {Nahum},
  \citenamefont {Vijay},\ and\ \citenamefont {Haah}}]{nahum2018operator}%
  \BibitemOpen
  \bibfield  {author} {\bibinfo {author} {\bibfnamefont {A.}~\bibnamefont
  {Nahum}}, \bibinfo {author} {\bibfnamefont {S.}~\bibnamefont {Vijay}},\ and\
  \bibinfo {author} {\bibfnamefont {J.}~\bibnamefont {Haah}},\ }\bibfield
  {title} {\bibinfo {title} {Operator spreading in random unitary circuits},\
  }\href {https://doi.org/10.1103/PhysRevX.8.021014} {\bibfield  {journal}
  {\bibinfo  {journal} {Phys. Rev. X}\ }\textbf {\bibinfo {volume} {8}},\
  \bibinfo {pages} {021014} (\bibinfo {year} {2018})}\BibitemShut {NoStop}%
\bibitem [{\citenamefont {Khemani}\ \emph {et~al.}(2018)\citenamefont
  {Khemani}, \citenamefont {Vishwanath},\ and\ \citenamefont
  {Huse}}]{khemani2018operator}%
  \BibitemOpen
  \bibfield  {author} {\bibinfo {author} {\bibfnamefont {V.}~\bibnamefont
  {Khemani}}, \bibinfo {author} {\bibfnamefont {A.}~\bibnamefont
  {Vishwanath}},\ and\ \bibinfo {author} {\bibfnamefont {D.~A.}\ \bibnamefont
  {Huse}},\ }\bibfield  {title} {\bibinfo {title} {Operator spreading and the
  emergence of dissipative hydrodynamics under unitary evolution with
  conservation laws},\ }\href {https://doi.org/10.1103/PhysRevX.8.031057}
  {\bibfield  {journal} {\bibinfo  {journal} {Phys. Rev. X}\ }\textbf {\bibinfo
  {volume} {8}},\ \bibinfo {pages} {031057} (\bibinfo {year}
  {2018})}\BibitemShut {NoStop}%
\bibitem [{\citenamefont {von Keyserlingk}\ \emph {et~al.}(2018)\citenamefont
  {von Keyserlingk}, \citenamefont {Rakovszky}, \citenamefont {Pollmann},\ and\
  \citenamefont {Sondhi}}]{vonKeyserlingk2018operator}%
  \BibitemOpen
  \bibfield  {author} {\bibinfo {author} {\bibfnamefont {C.~W.}\ \bibnamefont
  {von Keyserlingk}}, \bibinfo {author} {\bibfnamefont {T.}~\bibnamefont
  {Rakovszky}}, \bibinfo {author} {\bibfnamefont {F.}~\bibnamefont
  {Pollmann}},\ and\ \bibinfo {author} {\bibfnamefont {S.~L.}\ \bibnamefont
  {Sondhi}},\ }\bibfield  {title} {\bibinfo {title} {Operator hydrodynamics,
  otocs, and entanglement growth in systems without conservation laws},\ }\href
  {https://doi.org/10.1103/PhysRevX.8.021013} {\bibfield  {journal} {\bibinfo
  {journal} {Phys. Rev. X}\ }\textbf {\bibinfo {volume} {8}},\ \bibinfo {pages}
  {021013} (\bibinfo {year} {2018})}\BibitemShut {NoStop}%
\bibitem [{\citenamefont {Rakovszky}\ \emph {et~al.}(2018)\citenamefont
  {Rakovszky}, \citenamefont {Pollmann},\ and\ \citenamefont {von
  Keyserlingk}}]{diffusive2018rakovszky}%
  \BibitemOpen
  \bibfield  {author} {\bibinfo {author} {\bibfnamefont {T.}~\bibnamefont
  {Rakovszky}}, \bibinfo {author} {\bibfnamefont {F.}~\bibnamefont
  {Pollmann}},\ and\ \bibinfo {author} {\bibfnamefont {C.~W.}\ \bibnamefont
  {von Keyserlingk}},\ }\bibfield  {title} {\bibinfo {title} {Diffusive
  hydrodynamics of out-of-time-ordered correlators with charge conservation},\
  }\href {https://doi.org/10.1103/PhysRevX.8.031058} {\bibfield  {journal}
  {\bibinfo  {journal} {Phys. Rev. X}\ }\textbf {\bibinfo {volume} {8}},\
  \bibinfo {pages} {031058} (\bibinfo {year} {2018})}\BibitemShut {NoStop}%
\bibitem [{\citenamefont {Rakovszky}\ \emph {et~al.}(2019)\citenamefont
  {Rakovszky}, \citenamefont {Pollmann},\ and\ \citenamefont {von
  Keyserlingk}}]{Rakovszky2019}%
  \BibitemOpen
  \bibfield  {author} {\bibinfo {author} {\bibfnamefont {T.}~\bibnamefont
  {Rakovszky}}, \bibinfo {author} {\bibfnamefont {F.}~\bibnamefont
  {Pollmann}},\ and\ \bibinfo {author} {\bibfnamefont {C.~W.}\ \bibnamefont
  {von Keyserlingk}},\ }\bibfield  {title} {\bibinfo {title} {Sub-ballistic
  growth of r\'enyi entropies due to diffusion},\ }\href
  {https://doi.org/10.1103/PhysRevLett.122.250602} {\bibfield  {journal}
  {\bibinfo  {journal} {Phys. Rev. Lett.}\ }\textbf {\bibinfo {volume} {122}},\
  \bibinfo {pages} {250602} (\bibinfo {year} {2019})}\BibitemShut {NoStop}%
\bibitem [{\citenamefont {Gullans}\ and\ \citenamefont
  {Huse}(2019)}]{Gullans2019}%
  \BibitemOpen
  \bibfield  {author} {\bibinfo {author} {\bibfnamefont {M.~J.}\ \bibnamefont
  {Gullans}}\ and\ \bibinfo {author} {\bibfnamefont {D.~A.}\ \bibnamefont
  {Huse}},\ }\bibfield  {title} {\bibinfo {title} {Entanglement structure of
  current-driven diffusive fermion systems},\ }\href
  {https://doi.org/10.1103/physrevx.9.021007} {\bibfield  {journal} {\bibinfo
  {journal} {Phys. Rev. X}\ }\textbf {\bibinfo {volume} {9}},\ \bibinfo {pages}
  {021007} (\bibinfo {year} {2019})}\BibitemShut {NoStop}%
\bibitem [{\citenamefont {McCulloch}\ \emph {et~al.}(2023)\citenamefont
  {McCulloch}, \citenamefont {De~Nardis}, \citenamefont {Gopalakrishnan},\ and\
  \citenamefont {Vasseur}}]{FCSMFTDeNardis}%
  \BibitemOpen
  \bibfield  {author} {\bibinfo {author} {\bibfnamefont {E.}~\bibnamefont
  {McCulloch}}, \bibinfo {author} {\bibfnamefont {J.}~\bibnamefont
  {De~Nardis}}, \bibinfo {author} {\bibfnamefont {S.}~\bibnamefont
  {Gopalakrishnan}},\ and\ \bibinfo {author} {\bibfnamefont {R.}~\bibnamefont
  {Vasseur}},\ }\bibfield  {title} {\bibinfo {title} {Full counting statistics
  of charge in chaotic many-body quantum systems},\ }\href
  {https://doi.org/10.1103/PhysRevLett.131.210402} {\bibfield  {journal}
  {\bibinfo  {journal} {Phys. Rev. Lett.}\ }\textbf {\bibinfo {volume} {131}},\
  \bibinfo {pages} {210402} (\bibinfo {year} {2023})}\BibitemShut {NoStop}%
\bibitem [{\citenamefont {Bauer}\ \emph {et~al.}(2017)\citenamefont {Bauer},
  \citenamefont {Bernard},\ and\ \citenamefont {Jin}}]{Bauer2017}%
  \BibitemOpen
  \bibfield  {author} {\bibinfo {author} {\bibfnamefont {M.}~\bibnamefont
  {Bauer}}, \bibinfo {author} {\bibfnamefont {D.}~\bibnamefont {Bernard}},\
  and\ \bibinfo {author} {\bibfnamefont {T.}~\bibnamefont {Jin}},\ }\bibfield
  {title} {\bibinfo {title} {{Stochastic dissipative quantum spin chains (I) :
  Quantum fluctuating discrete hydrodynamics}},\ }\href
  {https://doi.org/https://doi.org/10.21468/SciPostPhys.3.5.033} {\bibfield
  {journal} {\bibinfo  {journal} {SciPost Phys.}\ }\textbf {\bibinfo {volume}
  {3}},\ \bibinfo {pages} {033} (\bibinfo {year} {2017})}\BibitemShut {NoStop}%
\bibitem [{\citenamefont {Bauer}\ \emph {et~al.}(2019)\citenamefont {Bauer},
  \citenamefont {Bernard},\ and\ \citenamefont {Jin}}]{Bauer2019}%
  \BibitemOpen
  \bibfield  {author} {\bibinfo {author} {\bibfnamefont {M.}~\bibnamefont
  {Bauer}}, \bibinfo {author} {\bibfnamefont {D.}~\bibnamefont {Bernard}},\
  and\ \bibinfo {author} {\bibfnamefont {T.}~\bibnamefont {Jin}},\ }\bibfield
  {title} {\bibinfo {title} {Equilibrium fluctuations in maximally noisy
  extended quantum systems},\ }\href
  {https://doi.org/10.21468/scipostphys.6.4.045} {\bibfield  {journal}
  {\bibinfo  {journal} {SciPost Phys.}\ }\textbf {\bibinfo {volume} {6}},\
  \bibinfo {pages} {045} (\bibinfo {year} {2019})}\BibitemShut {NoStop}%
\bibitem [{\citenamefont {Bernard}\ \emph {et~al.}(2018)\citenamefont
  {Bernard}, \citenamefont {Jin},\ and\ \citenamefont
  {Shpielberg}}]{Bernard2018}%
  \BibitemOpen
  \bibfield  {author} {\bibinfo {author} {\bibfnamefont {D.}~\bibnamefont
  {Bernard}}, \bibinfo {author} {\bibfnamefont {T.}~\bibnamefont {Jin}},\ and\
  \bibinfo {author} {\bibfnamefont {O.}~\bibnamefont {Shpielberg}},\ }\bibfield
   {title} {\bibinfo {title} {Transport in quantum chains under strong
  monitoring},\ }\href {https://doi.org/10.1209/0295-5075/121/60006} {\bibfield
   {journal} {\bibinfo  {journal} {EPL}\ }\textbf {\bibinfo {volume} {121}},\
  \bibinfo {pages} {60006} (\bibinfo {year} {2018})}\BibitemShut {NoStop}%
\bibitem [{Ber(2019)}]{Bernard2019}%
  \BibitemOpen
  \bibfield  {title} {\bibinfo {title} {Open quantum symmetric simple exclusion
  process},\ }\href {https://doi.org/10.1103/PHYSREVLETT.123.080601} {\bibfield
   {journal} {\bibinfo  {journal} {Phys. Rev. Lett.}\ }\textbf {\bibinfo
  {volume} {123}},\ \bibinfo {pages} {080601} (\bibinfo {year}
  {2019})}\BibitemShut {NoStop}%
\bibitem [{\citenamefont {Bernard}\ and\ \citenamefont
  {Hruza}(2023)}]{Bernard2023}%
  \BibitemOpen
  \bibfield  {author} {\bibinfo {author} {\bibfnamefont {D.}~\bibnamefont
  {Bernard}}\ and\ \bibinfo {author} {\bibfnamefont {L.}~\bibnamefont
  {Hruza}},\ }\bibfield  {title} {\bibinfo {title} {Exact entanglement in the
  driven quantum symmetric simple exclusion process},\ }\href
  {https://doi.org/https://doi.org/10.21468/SciPostPhys.15.4.175} {\bibfield
  {journal} {\bibinfo  {journal} {SciPost Phys.}\ }\textbf {\bibinfo {volume}
  {15}},\ \bibinfo {pages} {175} (\bibinfo {year} {2023})}\BibitemShut
  {NoStop}%
\bibitem [{\citenamefont {Hruza}\ and\ \citenamefont
  {Bernard}(2023)}]{Hruza2023}%
  \BibitemOpen
  \bibfield  {author} {\bibinfo {author} {\bibfnamefont {L.}~\bibnamefont
  {Hruza}}\ and\ \bibinfo {author} {\bibfnamefont {D.}~\bibnamefont
  {Bernard}},\ }\bibfield  {title} {\bibinfo {title} {Coherent fluctuations in
  noisy mesoscopic systems, the open quantum ssep, and free probability},\
  }\href {https://doi.org/10.1103/PHYSREVX.13.011045/E011045_19.EPS/MEDIUM}
  {\bibfield  {journal} {\bibinfo  {journal} {Phys. Rev. X}\ }\textbf {\bibinfo
  {volume} {13}},\ \bibinfo {pages} {011045} (\bibinfo {year}
  {2023})}\BibitemShut {NoStop}%
\bibitem [{\citenamefont {Bernard}\ and\ \citenamefont
  {Hruza}(2024)}]{Bernard2024}%
  \BibitemOpen
  \bibfield  {author} {\bibinfo {author} {\bibfnamefont {D.}~\bibnamefont
  {Bernard}}\ and\ \bibinfo {author} {\bibfnamefont {L.}~\bibnamefont
  {Hruza}},\ }\bibfield  {title} {\bibinfo {title} {Structured random matrices
  and cyclic cumulants: A free probability approach},\ }\href
  {https://doi.org/10.1142/S201032632450014X} {\bibfield  {journal} {\bibinfo
  {journal} {Random Matrices Theory Appl.}\ }\textbf {\bibinfo {volume} {13}},\
  \bibinfo {pages} {2450014} (\bibinfo {year} {2024})}\BibitemShut {NoStop}%
\bibitem [{\citenamefont {Bauer}\ \emph {et~al.}(2023)\citenamefont {Bauer},
  \citenamefont {Bernard}, \citenamefont {Biane},\ and\ \citenamefont
  {Hruza}}]{Bauer2024}%
  \BibitemOpen
  \bibfield  {author} {\bibinfo {author} {\bibfnamefont {M.}~\bibnamefont
  {Bauer}}, \bibinfo {author} {\bibfnamefont {D.}~\bibnamefont {Bernard}},
  \bibinfo {author} {\bibfnamefont {P.}~\bibnamefont {Biane}},\ and\ \bibinfo
  {author} {\bibfnamefont {L.}~\bibnamefont {Hruza}},\ }\bibfield  {title}
  {\bibinfo {title} {Bernoulli variables, classical exclusion processes and
  free probability},\ }\href {https://doi.org/10.1007/s00023-023-01320-2}
  {\bibfield  {journal} {\bibinfo  {journal} {Ann. Henri Poincaré}\ }\textbf
  {\bibinfo {volume} {25}},\ \bibinfo {pages} {125–172} (\bibinfo {year}
  {2023})}\BibitemShut {NoStop}%
\bibitem [{\citenamefont {Bernard}\ \emph {et~al.}(2025)\citenamefont
  {Bernard}, \citenamefont {Jin}, \citenamefont {Scopa},\ and\ \citenamefont
  {Wei}}]{Bernard2025}%
  \BibitemOpen
  \bibfield  {author} {\bibinfo {author} {\bibfnamefont {D.}~\bibnamefont
  {Bernard}}, \bibinfo {author} {\bibfnamefont {T.}~\bibnamefont {Jin}},
  \bibinfo {author} {\bibfnamefont {S.}~\bibnamefont {Scopa}},\ and\ \bibinfo
  {author} {\bibfnamefont {S.}~\bibnamefont {Wei}},\ }\bibfield  {title}
  {\bibinfo {title} {Large deviations of density fluctuations in the
  boundary-driven quantum symmetric simple inclusion process},\ }\href
  {https://doi.org/10.1103/fs7f-z3k6} {\bibfield  {journal} {\bibinfo
  {journal} {Phys. Rev. E}\ }\textbf {\bibinfo {volume} {112}},\ \bibinfo
  {pages} {034106} (\bibinfo {year} {2025})}\BibitemShut {NoStop}%
\bibitem [{\citenamefont {Albert}\ \emph {et~al.}(2026)\citenamefont {Albert},
  \citenamefont {Bernard}, \citenamefont {Jin}, \citenamefont {Scopa},\ and\
  \citenamefont {Wei}}]{albert2026}%
  \BibitemOpen
  \bibfield  {author} {\bibinfo {author} {\bibfnamefont {M.}~\bibnamefont
  {Albert}}, \bibinfo {author} {\bibfnamefont {D.}~\bibnamefont {Bernard}},
  \bibinfo {author} {\bibfnamefont {T.}~\bibnamefont {Jin}}, \bibinfo {author}
  {\bibfnamefont {S.}~\bibnamefont {Scopa}},\ and\ \bibinfo {author}
  {\bibfnamefont {S.}~\bibnamefont {Wei}},\ }\href@noop {} {\bibinfo {title}
  {Universal classical and quantum fluctuations in the large deviations of
  current of noisy quantum systems: The case of qssep and qssip}} (\bibinfo
  {year} {2026}),\ \Eprint {https://arxiv.org/abs/arXiv:2601.16883}
  {arXiv:2601.16883} \BibitemShut {NoStop}%
\bibitem [{\citenamefont {Bernard}(2026)}]{Bernard2026-qssep-cont}%
  \BibitemOpen
  \bibfield  {author} {\bibinfo {author} {\bibfnamefont {D.}~\bibnamefont
  {Bernard}},\ }\href@noop {} {\bibinfo {title} {The quantum symmetric simple
  exclusion process in the continuum and free processes}} (\bibinfo {year}
  {2026}),\ \Eprint {https://arxiv.org/abs/arXiv:2602.16544} {arXiv:2602.16544}
  \BibitemShut {NoStop}%
\bibitem [{\citenamefont {Barraquand}\ and\ \citenamefont
  {Bernard}(2025)}]{Barraquand2026}%
  \BibitemOpen
  \bibfield  {author} {\bibinfo {author} {\bibfnamefont {G.}~\bibnamefont
  {Barraquand}}\ and\ \bibinfo {author} {\bibfnamefont {D.}~\bibnamefont
  {Bernard}},\ }\href@noop {} {\bibinfo {title} {Introduction to quantum
  exclusion processes}} (\bibinfo {year} {2025}),\ \Eprint
  {https://arxiv.org/abs/arXiv:2507.01570} {arXiv:2507.01570} \BibitemShut
  {NoStop}%
\bibitem [{\citenamefont {Costa}\ \emph {et~al.}(2025)\citenamefont {Costa},
  \citenamefont {Ribeiro},\ and\ \citenamefont {Luca}}]{costa2025}%
  \BibitemOpen
  \bibfield  {author} {\bibinfo {author} {\bibfnamefont {J.}~\bibnamefont
  {Costa}}, \bibinfo {author} {\bibfnamefont {P.}~\bibnamefont {Ribeiro}},\
  and\ \bibinfo {author} {\bibfnamefont {A.~D.}\ \bibnamefont {Luca}},\
  }\href@noop {} {\bibinfo {title} {Emergence of universality in transport of
  noisy free fermions}} (\bibinfo {year} {2025}),\ \Eprint
  {https://arxiv.org/abs/arXiv:2504.00188} {arXiv:2504.00188} \BibitemShut
  {NoStop}%
\bibitem [{\citenamefont {Alba}(2026)}]{alba2025nuqssep}%
  \BibitemOpen
  \bibfield  {author} {\bibinfo {author} {\bibfnamefont {V.}~\bibnamefont
  {Alba}},\ }\bibfield  {title} {\bibinfo {title} {Extension of quantum
  symmetric simple exclusion process model for entanglement spreading in
  stochastic diffusive quantum systems},\ }\href
  {https://doi.org/10.1103/kbgw-lrfs} {\bibfield  {journal} {\bibinfo
  {journal} {Phys. Rev. B}\ }\textbf {\bibinfo {volume} {113}},\ \bibinfo
  {pages} {075145} (\bibinfo {year} {2026})}\BibitemShut {NoStop}%
\bibitem [{\citenamefont {Russotto}\ \emph
  {et~al.}(2026{\natexlab{a}})\citenamefont {Russotto}, \citenamefont {Ares},
  \citenamefont {Calabrese},\ and\ \citenamefont {Alba}}]{russotto2026}%
  \BibitemOpen
  \bibfield  {author} {\bibinfo {author} {\bibfnamefont {A.}~\bibnamefont
  {Russotto}}, \bibinfo {author} {\bibfnamefont {F.}~\bibnamefont {Ares}},
  \bibinfo {author} {\bibfnamefont {P.}~\bibnamefont {Calabrese}},\ and\
  \bibinfo {author} {\bibfnamefont {V.}~\bibnamefont {Alba}},\ }\bibfield
  {title} {\bibinfo {title} {Dynamics of entanglement fluctuations and quantum
  mpemba effect in the $\nu=1$ qssep model},\ }\href
  {https://doi.org/10.1088/1742-5468/ae4bb9} {\bibfield  {journal} {\bibinfo
  {journal} {J. Stat. Mech.: Theory Exp.}\ }\textbf {\bibinfo {volume}
  {2026}}\bibinfo  {number} { (3)},\ \bibinfo {pages} {033103}}\BibitemShut
  {NoStop}%
\bibitem [{\citenamefont {Russotto}\ \emph
  {et~al.}(2026{\natexlab{b}})\citenamefont {Russotto}, \citenamefont {Ares},
  \citenamefont {Calabrese},\ and\ \citenamefont {Alba}}]{russotto2026b}%
  \BibitemOpen
\bibfield  {number} {  }\bibfield  {author} {\bibinfo {author} {\bibfnamefont
  {A.}~\bibnamefont {Russotto}}, \bibinfo {author} {\bibfnamefont
  {F.}~\bibnamefont {Ares}}, \bibinfo {author} {\bibfnamefont {P.}~\bibnamefont
  {Calabrese}},\ and\ \bibinfo {author} {\bibfnamefont {V.}~\bibnamefont
  {Alba}},\ }\bibfield  {title} {\bibinfo {title} {Inhomogeneous quenches and
  ghd in the $\nu=1$ qssep model},\ }\href
  {https://doi.org/10.1088/1742-5468/ae7423} {\bibfield  {journal} {\bibinfo
  {journal} {J. Stat. Mech.: Theory Exp.}\ }\textbf {\bibinfo {volume}
  {2026}}\bibinfo  {number} { (6)},\ \bibinfo {pages} {063102}}\BibitemShut
  {NoStop}%
\bibitem [{\citenamefont {Minoguchi}\ \emph {et~al.}(2025)\citenamefont
  {Minoguchi}, \citenamefont {Huber}, \citenamefont {Garbe}, \citenamefont
  {Gambassi},\ and\ \citenamefont {Rabl}}]{Minoguchi2023}%
  \BibitemOpen
\bibfield  {number} {  }\bibfield  {author} {\bibinfo {author} {\bibfnamefont
  {Y.}~\bibnamefont {Minoguchi}}, \bibinfo {author} {\bibfnamefont
  {J.}~\bibnamefont {Huber}}, \bibinfo {author} {\bibfnamefont
  {L.}~\bibnamefont {Garbe}}, \bibinfo {author} {\bibfnamefont
  {A.}~\bibnamefont {Gambassi}},\ and\ \bibinfo {author} {\bibfnamefont
  {P.}~\bibnamefont {Rabl}},\ }\bibfield  {title} {\bibinfo {title} {Unified
  interface model for dissipative transport of bosons and fermions},\ }\href
  {https://doi.org/10.1103/PhysRevLett.134.207102} {\bibfield  {journal}
  {\bibinfo  {journal} {Phys. Rev. Lett.}\ }\textbf {\bibinfo {volume} {134}},\
  \bibinfo {pages} {207102} (\bibinfo {year} {2025})}\BibitemShut {NoStop}%
\bibitem [{\citenamefont {Funaki}\ \emph {et~al.}(1991)\citenamefont {Funaki},
  \citenamefont {Handa},\ and\ \citenamefont {Uchiyama}}]{Funaki1991}%
  \BibitemOpen
  \bibfield  {author} {\bibinfo {author} {\bibfnamefont {T.}~\bibnamefont
  {Funaki}}, \bibinfo {author} {\bibfnamefont {K.}~\bibnamefont {Handa}},\ and\
  \bibinfo {author} {\bibfnamefont {K.}~\bibnamefont {Uchiyama}},\ }\bibfield
  {title} {\bibinfo {title} {Hydrodynamic limit of one-dimensional exclusion
  processes with speed change},\ }\href
  {https://doi.org/10.1214/aop/1176990543} {\bibfield  {journal} {\bibinfo
  {journal} {Ann. Probab.}\ }\textbf {\bibinfo {volume} {19}},\ \bibinfo
  {pages} {245} (\bibinfo {year} {1991})}\BibitemShut {NoStop}%
\bibitem [{\citenamefont {Gon\c{c}alves}\ \emph {et~al.}(2009)\citenamefont
  {Gon\c{c}alves}, \citenamefont {Landim},\ and\ \citenamefont
  {Toninelli}}]{Gonalves2009}%
  \BibitemOpen
  \bibfield  {author} {\bibinfo {author} {\bibfnamefont {P.}~\bibnamefont
  {Gon\c{c}alves}}, \bibinfo {author} {\bibfnamefont {C.}~\bibnamefont
  {Landim}},\ and\ \bibinfo {author} {\bibfnamefont {C.}~\bibnamefont
  {Toninelli}},\ }\bibfield  {title} {\bibinfo {title} {Hydrodynamic limit for
  a particle system with degenerate rates},\ }\href
  {https://doi.org/10.1214/09-aihp210} {\bibfield  {journal} {\bibinfo
  {journal} {Ann. Inst. Henri Poincaré Probab. Stat.}\ }\textbf {\bibinfo
  {volume} {45}},\ \bibinfo {pages} {887} (\bibinfo {year} {2009})}\BibitemShut
  {NoStop}%
\bibitem [{\citenamefont {Blondel}\ \emph {et~al.}(2020)\citenamefont
  {Blondel}, \citenamefont {Erignoux}, \citenamefont {Sasada},\ and\
  \citenamefont {Simon}}]{Blondel2020}%
  \BibitemOpen
  \bibfield  {author} {\bibinfo {author} {\bibfnamefont {O.}~\bibnamefont
  {Blondel}}, \bibinfo {author} {\bibfnamefont {C.}~\bibnamefont {Erignoux}},
  \bibinfo {author} {\bibfnamefont {M.}~\bibnamefont {Sasada}},\ and\ \bibinfo
  {author} {\bibfnamefont {M.}~\bibnamefont {Simon}},\ }\bibfield  {title}
  {\bibinfo {title} {Hydrodynamic limit for a facilitated exclusion process},\
  }\href {https://doi.org/10.1214/19-aihp977} {\bibfield  {journal} {\bibinfo
  {journal} {Ann. Inst. Henri Poincaré Probab. Stat.}\ }\textbf {\bibinfo
  {volume} {56}},\ \bibinfo {pages} {667} (\bibinfo {year} {2020})}\BibitemShut
  {NoStop}%
\bibitem [{\citenamefont {Erignoux}\ \emph {et~al.}(2024)\citenamefont
  {Erignoux}, \citenamefont {Simon},\ and\ \citenamefont
  {Zhao}}]{Erignoux2024}%
  \BibitemOpen
  \bibfield  {author} {\bibinfo {author} {\bibfnamefont {C.}~\bibnamefont
  {Erignoux}}, \bibinfo {author} {\bibfnamefont {M.}~\bibnamefont {Simon}},\
  and\ \bibinfo {author} {\bibfnamefont {L.}~\bibnamefont {Zhao}},\ }\bibfield
  {title} {\bibinfo {title} {Mapping hydrodynamics for the facilitated
  exclusion and zero-range processes},\ }\href
  {https://doi.org/10.1214/23-AAP1997} {\bibfield  {journal} {\bibinfo
  {journal} {Ann. Appl. Probab.}\ }\textbf {\bibinfo {volume} {34}},\ \bibinfo
  {pages} {1524} (\bibinfo {year} {2024})}\BibitemShut {NoStop}%
\bibitem [{\citenamefont {McCulloch}\ \emph
  {et~al.}(2026{\natexlab{a}})\citenamefont {McCulloch}, \citenamefont
  {Jacoby}, \citenamefont {von Keyserlingk},\ and\ \citenamefont
  {Gopalakrishnan}}]{McCulloch2026}%
  \BibitemOpen
  \bibfield  {author} {\bibinfo {author} {\bibfnamefont {E.}~\bibnamefont
  {McCulloch}}, \bibinfo {author} {\bibfnamefont {J.~A.}\ \bibnamefont
  {Jacoby}}, \bibinfo {author} {\bibfnamefont {C.}~\bibnamefont {von
  Keyserlingk}},\ and\ \bibinfo {author} {\bibfnamefont {S.}~\bibnamefont
  {Gopalakrishnan}},\ }\bibfield  {title} {\bibinfo {title} {Subexponential
  decay of local correlations from diffusion-limited dephasing},\ }\href
  {https://doi.org/10.1103/393g-z21y} {\bibfield  {journal} {\bibinfo
  {journal} {Phys. Rev. Lett.}\ }\textbf {\bibinfo {volume} {136}},\ \bibinfo
  {pages} {190403} (\bibinfo {year} {2026}{\natexlab{a}})}\BibitemShut
  {NoStop}%
\bibitem [{\citenamefont {McCulloch}\ \emph
  {et~al.}(2026{\natexlab{b}})\citenamefont {McCulloch}, \citenamefont
  {Jacoby},\ and\ \citenamefont {Gopalakrishnan}}]{McCulloch2026b}%
  \BibitemOpen
  \bibfield  {author} {\bibinfo {author} {\bibfnamefont {E.}~\bibnamefont
  {McCulloch}}, \bibinfo {author} {\bibfnamefont {J.~A.}\ \bibnamefont
  {Jacoby}},\ and\ \bibinfo {author} {\bibfnamefont {S.}~\bibnamefont
  {Gopalakrishnan}},\ }\href@noop {} {\bibinfo {title} {Long-lived local
  quantum coherences from hydrodynamic large deviations}} (\bibinfo {year}
  {2026}{\natexlab{b}}),\ \Eprint {https://arxiv.org/abs/arXiv:2604.27074}
  {arXiv:2604.27074} \BibitemShut {NoStop}%
\bibitem [{\citenamefont {Kob}\ and\ \citenamefont {Andersen}(1993)}]{Kob1993}%
  \BibitemOpen
  \bibfield  {author} {\bibinfo {author} {\bibfnamefont {W.}~\bibnamefont
  {Kob}}\ and\ \bibinfo {author} {\bibfnamefont {H.~C.}\ \bibnamefont
  {Andersen}},\ }\bibfield  {title} {\bibinfo {title} {Kinetic lattice-gas
  model of cage effects in high-density liquids and a test of mode-coupling
  theory of the ideal-glass transition},\ }\href
  {https://doi.org/10.1103/physreve.48.4364} {\bibfield  {journal} {\bibinfo
  {journal} {Phys. Rev. E}\ }\textbf {\bibinfo {volume} {48}},\ \bibinfo
  {pages} {4364–4377} (\bibinfo {year} {1993})}\BibitemShut {NoStop}%
\bibitem [{\citenamefont {Rossi}\ \emph {et~al.}(2000)\citenamefont {Rossi},
  \citenamefont {Pastor-Satorras},\ and\ \citenamefont
  {Vespignani}}]{Rossi2000}%
  \BibitemOpen
  \bibfield  {author} {\bibinfo {author} {\bibfnamefont {M.}~\bibnamefont
  {Rossi}}, \bibinfo {author} {\bibfnamefont {R.}~\bibnamefont
  {Pastor-Satorras}},\ and\ \bibinfo {author} {\bibfnamefont {A.}~\bibnamefont
  {Vespignani}},\ }\bibfield  {title} {\bibinfo {title} {Universality class of
  absorbing phase transitions with a conserved field},\ }\href
  {https://doi.org/10.1103/physrevlett.85.1803} {\bibfield  {journal} {\bibinfo
   {journal} {Phys. Rev. Lett.}\ }\textbf {\bibinfo {volume} {85}},\ \bibinfo
  {pages} {1803–1806} (\bibinfo {year} {2000})}\BibitemShut {NoStop}%
\bibitem [{\citenamefont {de~Oliveira}(2005)}]{deOliveira2005}%
  \BibitemOpen
  \bibfield  {author} {\bibinfo {author} {\bibfnamefont {M.~J.}\ \bibnamefont
  {de~Oliveira}},\ }\bibfield  {title} {\bibinfo {title} {Conserved lattice gas
  model with infinitely many absorbing states in one dimension},\ }\href
  {https://doi.org/10.1103/physreve.71.016112} {\bibfield  {journal} {\bibinfo
  {journal} {Phys. Rev. E}\ }\textbf {\bibinfo {volume} {71}},\ \bibinfo
  {pages} {016112} (\bibinfo {year} {2005})}\BibitemShut {NoStop}%
\bibitem [{\citenamefont {Gabel}\ \emph {et~al.}(2010)\citenamefont {Gabel},
  \citenamefont {Krapivsky},\ and\ \citenamefont {Redner}}]{Gabel2010}%
  \BibitemOpen
  \bibfield  {author} {\bibinfo {author} {\bibfnamefont {A.}~\bibnamefont
  {Gabel}}, \bibinfo {author} {\bibfnamefont {P.~L.}\ \bibnamefont
  {Krapivsky}},\ and\ \bibinfo {author} {\bibfnamefont {S.}~\bibnamefont
  {Redner}},\ }\bibfield  {title} {\bibinfo {title} {Facilitated asymmetric
  exclusion},\ }\href {https://doi.org/10.1103/physrevlett.105.210603}
  {\bibfield  {journal} {\bibinfo  {journal} {Phys. Rev. Lett.}\ }\textbf
  {\bibinfo {volume} {105}},\ \bibinfo {pages} {210603} (\bibinfo {year}
  {2010})}\BibitemShut {NoStop}%
\bibitem [{\citenamefont {Baik}\ \emph {et~al.}(2018)\citenamefont {Baik},
  \citenamefont {Barraquand}, \citenamefont {Corwin},\ and\ \citenamefont
  {Suidan}}]{Baik2018}%
  \BibitemOpen
  \bibfield  {author} {\bibinfo {author} {\bibfnamefont {J.}~\bibnamefont
  {Baik}}, \bibinfo {author} {\bibfnamefont {G.}~\bibnamefont {Barraquand}},
  \bibinfo {author} {\bibfnamefont {I.}~\bibnamefont {Corwin}},\ and\ \bibinfo
  {author} {\bibfnamefont {T.}~\bibnamefont {Suidan}},\ }\bibinfo {title}
  {Facilitated exclusion process},\ in\ \href
  {https://doi.org/10.1007/978-3-030-01593-0_1} {\emph {\bibinfo {booktitle}
  {Computation and Combinatorics in Dynamics, Stochastics and Control}}}\
  (\bibinfo  {publisher} {Springer International Publishing},\ \bibinfo {year}
  {2018})\ p.\ \bibinfo {pages} {1–35}\BibitemShut {NoStop}%
\bibitem [{\citenamefont {Goldstein}\ \emph {et~al.}(2019)\citenamefont
  {Goldstein}, \citenamefont {Lebowitz},\ and\ \citenamefont
  {Speer}}]{Goldstein2019}%
  \BibitemOpen
  \bibfield  {author} {\bibinfo {author} {\bibfnamefont {S.}~\bibnamefont
  {Goldstein}}, \bibinfo {author} {\bibfnamefont {J.~L.}\ \bibnamefont
  {Lebowitz}},\ and\ \bibinfo {author} {\bibfnamefont {E.~R.}\ \bibnamefont
  {Speer}},\ }\bibfield  {title} {\bibinfo {title} {Exact solution of the
  facilitated totally asymmetric simple exclusion process},\ }\href
  {https://doi.org/10.1088/1742-5468/ab363f} {\bibfield  {journal} {\bibinfo
  {journal} {J. Stat. Mech.: Theory Exp.}\ }\textbf {\bibinfo {volume}
  {2019}}\bibinfo  {number} { (12)},\ \bibinfo {pages} {123202}}\BibitemShut
  {NoStop}%
\bibitem [{\citenamefont {Ayyer}\ \emph {et~al.}(2023)\citenamefont {Ayyer},
  \citenamefont {Goldstein}, \citenamefont {Lebowitz},\ and\ \citenamefont
  {Speer}}]{Ayyer2023}%
  \BibitemOpen
\bibfield  {number} {  }\bibfield  {author} {\bibinfo {author} {\bibfnamefont
  {A.}~\bibnamefont {Ayyer}}, \bibinfo {author} {\bibfnamefont
  {S.}~\bibnamefont {Goldstein}}, \bibinfo {author} {\bibfnamefont {J.~L.}\
  \bibnamefont {Lebowitz}},\ and\ \bibinfo {author} {\bibfnamefont {E.~R.}\
  \bibnamefont {Speer}},\ }\bibfield  {title} {\bibinfo {title} {Stationary
  states of the one-dimensional facilitated asymmetric exclusion process},\
  }\href {https://doi.org/10.1214/22-aihp1264} {\bibfield  {journal} {\bibinfo
  {journal} {Ann. Inst. Henri Poincaré Probab. Stat.}\ }\textbf {\bibinfo
  {volume} {59}},\ \bibinfo {pages} {726} (\bibinfo {year} {2023})}\BibitemShut
  {NoStop}%
\bibitem [{\citenamefont {Barraquand}\ \emph {et~al.}(2025)\citenamefont
  {Barraquand}, \citenamefont {Blondel},\ and\ \citenamefont
  {Simon}}]{Barraquand2025}%
  \BibitemOpen
  \bibfield  {author} {\bibinfo {author} {\bibfnamefont {G.}~\bibnamefont
  {Barraquand}}, \bibinfo {author} {\bibfnamefont {O.}~\bibnamefont
  {Blondel}},\ and\ \bibinfo {author} {\bibfnamefont {M.}~\bibnamefont
  {Simon}},\ }\bibfield  {title} {\bibinfo {title} {Weakly asymmetric
  facilitated exclusion process},\ }\href {https://doi.org/10.1214/25-ejp1390}
  {\bibfield  {journal} {\bibinfo  {journal} {Electron. J. Probab.}\ }\textbf
  {\bibinfo {volume} {30}},\ \bibinfo {pages} {128} (\bibinfo {year}
  {2025})}\BibitemShut {NoStop}%
\bibitem [{\citenamefont {Da~Cunha}\ \emph {et~al.}(2026)\citenamefont
  {Da~Cunha}, \citenamefont {Erignoux},\ and\ \citenamefont
  {Simon}}]{DaCunha2026}%
  \BibitemOpen
  \bibfield  {author} {\bibinfo {author} {\bibfnamefont {H.}~\bibnamefont
  {Da~Cunha}}, \bibinfo {author} {\bibfnamefont {C.}~\bibnamefont {Erignoux}},\
  and\ \bibinfo {author} {\bibfnamefont {M.}~\bibnamefont {Simon}},\ }\bibfield
   {title} {\bibinfo {title} {Hydrodynamic limit for an open facilitated
  exclusion process with slow and fast boundaries},\ }\href
  {https://doi.org/10.1007/s00220-025-05550-9} {\bibfield  {journal} {\bibinfo
  {journal} {Commun. Math. Phys.}\ }\textbf {\bibinfo {volume} {407}},\
  \bibinfo {pages} {50} (\bibinfo {year} {2026})}\BibitemShut {NoStop}%
\bibitem [{\citenamefont {Olmos}\ \emph {et~al.}(2012)\citenamefont {Olmos},
  \citenamefont {Lesanovsky},\ and\ \citenamefont {Garrahan}}]{Olmos2012}%
  \BibitemOpen
  \bibfield  {author} {\bibinfo {author} {\bibfnamefont {B.}~\bibnamefont
  {Olmos}}, \bibinfo {author} {\bibfnamefont {I.}~\bibnamefont {Lesanovsky}},\
  and\ \bibinfo {author} {\bibfnamefont {J.~P.}\ \bibnamefont {Garrahan}},\
  }\bibfield  {title} {\bibinfo {title} {Facilitated spin models of dissipative
  quantum glasses},\ }\href {https://doi.org/10.1103/physrevlett.109.020403}
  {\bibfield  {journal} {\bibinfo  {journal} {Phys. Rev. Lett.}\ }\textbf
  {\bibinfo {volume} {109}},\ \bibinfo {pages} {020403} (\bibinfo {year}
  {2012})}\BibitemShut {NoStop}%
\bibitem [{\citenamefont {Olmos}\ \emph {et~al.}(2014)\citenamefont {Olmos},
  \citenamefont {Lesanovsky},\ and\ \citenamefont {Garrahan}}]{Olmos2014}%
  \BibitemOpen
  \bibfield  {author} {\bibinfo {author} {\bibfnamefont {B.}~\bibnamefont
  {Olmos}}, \bibinfo {author} {\bibfnamefont {I.}~\bibnamefont {Lesanovsky}},\
  and\ \bibinfo {author} {\bibfnamefont {J.~P.}\ \bibnamefont {Garrahan}},\
  }\bibfield  {title} {\bibinfo {title} {Out-of-equilibrium evolution of
  kinetically constrained many-body quantum systems under purely dissipative
  dynamics},\ }\href {https://doi.org/10.1103/physreve.90.042147} {\bibfield
  {journal} {\bibinfo  {journal} {Phys. Rev. E}\ }\textbf {\bibinfo {volume}
  {90}},\ \bibinfo {pages} {042147} (\bibinfo {year} {2014})}\BibitemShut
  {NoStop}%
\bibitem [{\citenamefont {Pancotti}\ \emph {et~al.}(2020)\citenamefont
  {Pancotti}, \citenamefont {Giudice}, \citenamefont {Cirac}, \citenamefont
  {Garrahan},\ and\ \citenamefont {Bañuls}}]{Pancotti2020}%
  \BibitemOpen
  \bibfield  {author} {\bibinfo {author} {\bibfnamefont {N.}~\bibnamefont
  {Pancotti}}, \bibinfo {author} {\bibfnamefont {G.}~\bibnamefont {Giudice}},
  \bibinfo {author} {\bibfnamefont {J.~I.}\ \bibnamefont {Cirac}}, \bibinfo
  {author} {\bibfnamefont {J.~P.}\ \bibnamefont {Garrahan}},\ and\ \bibinfo
  {author} {\bibfnamefont {M.~C.}\ \bibnamefont {Bañuls}},\ }\bibfield
  {title} {\bibinfo {title} {Quantum east model: Localization, nonthermal
  eigenstates, and slow dynamics},\ }\href
  {https://doi.org/10.1103/physrevx.10.021051} {\bibfield  {journal} {\bibinfo
  {journal} {Phys. Rev. X}\ }\textbf {\bibinfo {volume} {10}},\ \bibinfo
  {pages} {021051} (\bibinfo {year} {2020})}\BibitemShut {NoStop}%
\bibitem [{\citenamefont {Rose}\ \emph {et~al.}(2022)\citenamefont {Rose},
  \citenamefont {Macieszczak}, \citenamefont {Lesanovsky},\ and\ \citenamefont
  {Garrahan}}]{Rose2022}%
  \BibitemOpen
  \bibfield  {author} {\bibinfo {author} {\bibfnamefont {D.~C.}\ \bibnamefont
  {Rose}}, \bibinfo {author} {\bibfnamefont {K.}~\bibnamefont {Macieszczak}},
  \bibinfo {author} {\bibfnamefont {I.}~\bibnamefont {Lesanovsky}},\ and\
  \bibinfo {author} {\bibfnamefont {J.~P.}\ \bibnamefont {Garrahan}},\
  }\bibfield  {title} {\bibinfo {title} {Hierarchical classical metastability
  in an open quantum east model},\ }\href
  {https://doi.org/10.1103/physreve.105.044121} {\bibfield  {journal} {\bibinfo
   {journal} {Phys. Rev. E}\ }\textbf {\bibinfo {volume} {105}},\ \bibinfo
  {pages} {044121} (\bibinfo {year} {2022})}\BibitemShut {NoStop}%
\bibitem [{\citenamefont {Brighi}\ \emph {et~al.}(2023)\citenamefont {Brighi},
  \citenamefont {Ljubotina},\ and\ \citenamefont {Serbyn}}]{Brighi2023}%
  \BibitemOpen
  \bibfield  {author} {\bibinfo {author} {\bibfnamefont {P.}~\bibnamefont
  {Brighi}}, \bibinfo {author} {\bibfnamefont {M.}~\bibnamefont {Ljubotina}},\
  and\ \bibinfo {author} {\bibfnamefont {M.}~\bibnamefont {Serbyn}},\
  }\bibfield  {title} {\bibinfo {title} {Hilbert space fragmentation and slow
  dynamics in particle-conserving quantum east models},\ }\href
  {https://doi.org/10.21468/scipostphys.15.3.093} {\bibfield  {journal}
  {\bibinfo  {journal} {SciPost Phys.}\ }\textbf {\bibinfo {volume} {15}},\
  \bibinfo {pages} {093} (\bibinfo {year} {2023})}\BibitemShut {NoStop}%
\bibitem [{\citenamefont {Bertini}\ \emph {et~al.}(2024)\citenamefont
  {Bertini}, \citenamefont {De~Fazio}, \citenamefont {Garrahan},\ and\
  \citenamefont {Klobas}}]{Bertini2024}%
  \BibitemOpen
  \bibfield  {author} {\bibinfo {author} {\bibfnamefont {B.}~\bibnamefont
  {Bertini}}, \bibinfo {author} {\bibfnamefont {C.}~\bibnamefont {De~Fazio}},
  \bibinfo {author} {\bibfnamefont {J.~P.}\ \bibnamefont {Garrahan}},\ and\
  \bibinfo {author} {\bibfnamefont {K.}~\bibnamefont {Klobas}},\ }\bibfield
  {title} {\bibinfo {title} {Exact quench dynamics of the floquet quantum east
  model at the deterministic point},\ }\href
  {https://doi.org/10.1103/physrevlett.132.120402} {\bibfield  {journal}
  {\bibinfo  {journal} {Phys. Rev. Lett.}\ }\textbf {\bibinfo {volume} {132}},\
  \bibinfo {pages} {120402} (\bibinfo {year} {2024})}\BibitemShut {NoStop}%
\bibitem [{\citenamefont {Zadnik}\ and\ \citenamefont
  {Fagotti}(2021)}]{Zadnik2021}%
  \BibitemOpen
  \bibfield  {author} {\bibinfo {author} {\bibfnamefont {L.}~\bibnamefont
  {Zadnik}}\ and\ \bibinfo {author} {\bibfnamefont {M.}~\bibnamefont
  {Fagotti}},\ }\bibfield  {title} {\bibinfo {title} {The folded spin-1/2 xxz
  model: I. diagonalisation, jamming, and ground state properties},\ }\href
  {https://doi.org/10.21468/scipostphyscore.4.2.010} {\bibfield  {journal}
  {\bibinfo  {journal} {SciPost Phys. Core}\ }\textbf {\bibinfo {volume} {4}},\
  \bibinfo {pages} {010} (\bibinfo {year} {2021})}\BibitemShut {NoStop}%
\bibitem [{\citenamefont {Zadnik}\ \emph {et~al.}(2021)\citenamefont {Zadnik},
  \citenamefont {Bidzhiev},\ and\ \citenamefont {Fagotti}}]{Zadnik2021b}%
  \BibitemOpen
  \bibfield  {author} {\bibinfo {author} {\bibfnamefont {L.}~\bibnamefont
  {Zadnik}}, \bibinfo {author} {\bibfnamefont {K.}~\bibnamefont {Bidzhiev}},\
  and\ \bibinfo {author} {\bibfnamefont {M.}~\bibnamefont {Fagotti}},\
  }\bibfield  {title} {\bibinfo {title} {The folded spin-1/2 xxz model: Ii.
  thermodynamics and hydrodynamics with a minimal set of charges},\ }\href
  {https://doi.org/10.21468/scipostphys.10.5.099} {\bibfield  {journal}
  {\bibinfo  {journal} {SciPost Phys.}\ }\textbf {\bibinfo {volume} {10}},\
  \bibinfo {pages} {099} (\bibinfo {year} {2021})}\BibitemShut {NoStop}%
\bibitem [{\citenamefont {Pozsgay}\ \emph {et~al.}(2021)\citenamefont
  {Pozsgay}, \citenamefont {Gombor}, \citenamefont {Hutsalyuk}, \citenamefont
  {Jiang}, \citenamefont {Pristyák},\ and\ \citenamefont
  {Vernier}}]{Pozsgay2021}%
  \BibitemOpen
  \bibfield  {author} {\bibinfo {author} {\bibfnamefont {B.}~\bibnamefont
  {Pozsgay}}, \bibinfo {author} {\bibfnamefont {T.}~\bibnamefont {Gombor}},
  \bibinfo {author} {\bibfnamefont {A.}~\bibnamefont {Hutsalyuk}}, \bibinfo
  {author} {\bibfnamefont {Y.}~\bibnamefont {Jiang}}, \bibinfo {author}
  {\bibfnamefont {L.}~\bibnamefont {Pristyák}},\ and\ \bibinfo {author}
  {\bibfnamefont {E.}~\bibnamefont {Vernier}},\ }\bibfield  {title} {\bibinfo
  {title} {Integrable spin chain with hilbert space fragmentation and solvable
  real-time dynamics},\ }\href {https://doi.org/10.1103/physreve.104.044106}
  {\bibfield  {journal} {\bibinfo  {journal} {Phys. Rev. E}\ }\textbf {\bibinfo
  {volume} {104}},\ \bibinfo {pages} {044106} (\bibinfo {year}
  {2021})}\BibitemShut {NoStop}%
\bibitem [{\citenamefont {Cardy}(1996)}]{Cardy1996}%
  \BibitemOpen
  \bibfield  {author} {\bibinfo {author} {\bibfnamefont {J.}~\bibnamefont
  {Cardy}},\ }\href {https://doi.org/10.1017/cbo9781316036440} {\emph {\bibinfo
  {title} {Scaling and Renormalization in Statistical Physics}}}\ (\bibinfo
  {publisher} {Cambridge University Press},\ \bibinfo {year}
  {1996})\BibitemShut {NoStop}%
\bibitem [{\citenamefont {Christopoulos}\ \emph {et~al.}(2026)\citenamefont
  {Christopoulos}, \citenamefont {Costa}, \citenamefont {Bernard},
  \citenamefont {De~Nardis}, \citenamefont {Jin}, \citenamefont
  {Lenar\v{c}i\v{c}},\ and\ \citenamefont {Scopa}}]{noisyXXZ-2026}%
  \BibitemOpen
  \bibfield  {author} {\bibinfo {author} {\bibfnamefont {A.}~\bibnamefont
  {Christopoulos}}, \bibinfo {author} {\bibfnamefont {J.}~\bibnamefont
  {Costa}}, \bibinfo {author} {\bibfnamefont {D.}~\bibnamefont {Bernard}},
  \bibinfo {author} {\bibfnamefont {J.}~\bibnamefont {De~Nardis}}, \bibinfo
  {author} {\bibfnamefont {T.}~\bibnamefont {Jin}}, \bibinfo {author}
  {\bibfnamefont {Z.}~\bibnamefont {Lenar\v{c}i\v{c}}},\ and\ \bibinfo {author}
  {\bibfnamefont {S.}~\bibnamefont {Scopa}},\ }\href@noop {} {\bibinfo {title}
  {Non-equilibrium steady state of the quantum noisy xxz spin chain via
  macroscopic fluctuation theory~--~{\it to appear}}} (\bibinfo {year}
  {2026})\BibitemShut {NoStop}%
\end{thebibliography}%
%
%
\clearpage
\onecolumngrid
\newpage

\setcounter{equation}{0}  
\setcounter{figure}{0}
\setcounter{page}{1}
\setcounter{section}{0}    
\renewcommand\thesection{\arabic{section}}    
\renewcommand\thesubsection{\arabic{subsection}}    
\renewcommand{\thetable}{S\arabic{table}}
\renewcommand{\theequation}{S\arabic{equation}}
\renewcommand{\thefigure}{S\arabic{figure}}
\setcounter{secnumdepth}{2}  
\makeatletter
\newcommand{\smtableofcontents}{%
  \section*{Contents}%
  \@starttoc{smtoc}%
}
\newcommand{\smsection}[1]{%
  \section{#1}%
  \addcontentsline{smtoc}{section}{\protect\numberline{\thesection}#1}%
}
\newcommand{\smsubsection}[1]{%
  \subsection{#1}%
  \addcontentsline{smtoc}{subsection}{\protect\numberline{\thesubsection}#1}%
}
\makeatother

\begin{center}
\hypertarget{SM}{\Large \textbf{Supplementary Material}}\\ \ \\
{\large \textbf{\titleinfo}}
\ \\ \ 
\end{center}
This Supplementary Material contains the derivation and the generalization to arbitrary replicas of the results presented in the main text, preceded by a brief explanation of the numerical implementation used to obtain the data in Fig.~\ref{fig:num} of the main text. Below, we shall adopt the notation introduced in the main text and in the \hyperlink{EM}{End Matter}.

\bigskip
{\smtableofcontents}
\hrulefill

\smsection{Numerical implementation}
\label{sec:sm-numerics}
In this section, we describe the numerical method used to generate the microscopic data shown in Fig.~\ref{fig:num} of the main text. Throughout the Letter, we consider a chain of $L\equiv N+1$ sites. The Hilbert space is then the full fermionic Fock space
\[
\mathscr{H}_L=\mathrm{span}\{|n_0,\dots,n_{N}\rangle,\; n_i=0,1\},
\qquad
\dim\mathscr{H}_L=2^L.
\]
For each stochastic realization, we evolve the full many-body density matrix $\hat\rho_t\in{\rm End}(\mathscr{H}_L)$, i.e. a $2^L\times 2^L$ matrix in the Fock space. \\

For each trajectory and each time step, two independent real Gaussian random variables $\eta_k$ and $\eta'_k$ are generated on each link, with
\be
\E[\eta_k]=\E[\eta'_k]=0,\qquad
\E[\eta_k\eta_m]=\E[\eta'_k\eta'_m]=\delta_{km},
\qquad
\E[\eta_k\eta'_m]=0.
\ee
The complex Brownian increments are then discretized as
\be
dW^k_t=\sqrt{\frac{dt}{2}}(\eta_k+\I\eta'_k); \quad d\overline{W}^k_t=\sqrt{\frac{dt}{2}}(\eta_k-\I\eta'_k),\qquad  \E[dW^k_t\,d\overline W^m_t]=dt\,\delta_{km}.
\ee
The stochastic Hamiltonian increment is (cf.~Eq.~\eqref{eq:dH-iqsep} of the main text)
\[
d\hat H_t= \sum_{k=0}^{N-1} \hat P_k \left( \hat\ell_k\,dW^k_t +\hat\ell^\dagger_k \, d\overline W^k_t\right).
\]
In the numerical implementation, the interaction is modeled as in Eq.~\eqref{eq:short-range-dressing} of the main text. In particular, denoting with $\mathbf{n}=(n_0,\dots,n_N)$ a configuration of the chain,
\be
P_k(\mathbf n)=\sqrt{D_0}\left[1+\lambda\Big(n_{k-1}(1-n_{k+2})+n_{k+2}(1-n_{k-1})\Big)\right].
\ee
For each Fock configuration $\mathbf{n}$, the code precomputes the allowed moves at each link, together with the corresponding value of $P_k(\mathbf n)$. This avoids constructing dense hopping operators at each step during the time evolution. The stochastic update is then implemented as
\be\label{eq:SM-unitary-step}
\hat\rho_t \mapsto
e^{-\I d\hat H_t}\hat\rho_t\,e^{\I d\hat H_t}.
\ee
The action of the matrix exponentials $e^{\pm\I d\hat H_t}$ is evaluated by sparse multiplications, so that the dense exponential matrix is never formed explicitly. Boundary driving is included through the standard Lindblad operators e.g. of Refs.~\cite{Bernard2019,Hruza2023,Bernard2025}, 
\be
{\cal L}_\text{bdry}(\hat\rho_t)= \sum_{p\in\{0,N\}}\sum_{\sigma=\pm}\left[(\hat{L}_p^\sigma) \hat\rho_t (\hat{L}_p^\sigma)^\dagger-\frac12\{(\hat L_p^\sigma)^\dagger(\hat L_p^\sigma),\hat \rho_t\}\right],
\ee
with jump operators
\be
\hat L_0^+=\sqrt{\alpha_0}\,\hat c^\dagger_0,\qquad
\hat L_0^-=\sqrt{\beta_0}\,\hat c_0,
\qquad
\hat L_N^+=\sqrt{\alpha_N}\,\hat c^\dagger_{N},\qquad
\hat L_N^-=\sqrt{\beta_N}\,\hat c_{N}.
\ee
In the simulations shown in the main text we take, $\alpha_0=1$, $\beta_0=0$, and $\alpha_N=0$, $\beta_N=1$, corresponding to boundary densities $n_a=1$ (left) and $n_b=0$ (right) in the thermodynamic limit.\\

For the boundary numerical evolution, the density matrix is vectorized. We use the convention,
\be
|\rho\rangle\rangle := \mathrm{vec}(\hat\rho), \quad\text{such that}\quad |\rho\rangle\rangle_{a+b\,2^L}= \rho_{ab},
\ee
and the vectorized state has dimension $4^L$. With this convention, left and right multiplication are represented as
\[
\mathrm{vec}(A\rho B)
=
(B^{\rm T}\otimes A)\,\mathrm{vec}(\rho).
\]
The deterministic boundary Lindbladian is then stored as a sparse superoperator acting on $|\rho\rangle\rangle$. Its action is implemented as,
\be
|\hat\rho_{t+dt}\rangle\rangle =e^{dt{\cal L}_\text{bdry}} \mathrm{vec}\Big( e^{-\I d\hat H_t}\ \hat\rho_t \ e^{\I d\hat H_t}\Big).
\ee
Therefore, the strategy for the time evolution goes as follows. At each time step, we apply the unitary update~\eqref{eq:SM-unitary-step} keeping the density matrix form. The state is then vectorized and evolved with ${\cal L}_\text{bdry}$. The vectorized state is then reshaped back into a $2^L\times2^L$ matrix, again unitarily evolved, and then vectorized again. This strategy avoids constructing the full bulk superoperator at each time step.\\

The initial condition is the infinite-temperature density matrix, $\hat\rho_0=\text{Id}_{2^L\times 2^L}/(2^L)$. Each trajectory is evolved up to a diffusive time $t_\text{fin}\sim N^2$, with time step $dt$. At the final time, for each trajectory $w$, we compute the density profile and the coherences 
\be
n_i^{(w)}={\rm tr}\big(\hat\rho_{t_\text{fin}}^{(w)}\hat n_i\big); \qquad G_{ij}^{(w)}={\rm tr}\big(\hat\rho_{t_\text{fin}}^{(w)}\hat X_{ij}\big).
\ee
 The numerical estimate of the averaged density and two-replica coherence loop is then obtained as
\be
n^\text{num}(i;t_\text{fin})=\overline{n_i^{(w)}}; \qquad   g^\text{num}_2(i,j; t_\text{fin})= N\big(\overline{G_{ij}^{(w)}G_{ji}^{(w)}}-\overline{G_{ij}^{(w)}}\ \overline{G_{ji}^{(w)}}\big).
\ee
Here, the overline denotes the statistical average over stochastic realizations, i.e. $\overline{f}:=\tfrac{1}{N_\text{traj}}\sum_{w=1}^{N_\text{traj}} f^{(w)}$.
\smsection{Single-replica hydrodynamics}
We start from Eq.~\eqref{eq:single-rep-hydro} of the \hyperlink{EM}{End Matter},
\be\label{eq:single-rep-hydro-sm}
\frac{d}{dt}\avg{\hat O_m}_t =\sum_{k=0}^{N-1} \avg{ \hat P_k {\cal L}_k^\text{qssep}(\hat O_m )\hat P_k }_t,
\ee
where $\hat O_m= \hat n_{i_1}\dots \hat n_{i_m}$ is a string of density operators. We then evaluate,
\begin{align}\label{eq:evaluate-Lqseep-density}
&{\cal L}^{\rm qssep}_k( \hat n_{i})=\left(\delta_{k,i}-\delta_{k,i-1}\right)\left(\hat n_{k+1}-\hat n_k\right);\nn
&{\cal L}^{\rm qssep}_k( \hat n_{i}\hat n_j )={\cal L}^{\rm qssep}_k( \hat n_{i})\hat n_j + \hat n_i {\cal L}^{\rm qssep}_k( \hat n_{j})-\mathbf{1}_{\{i,j\}=\{k,k+1\}} \hat\Delta_k;\nn
&\vdots\\
&{\cal L}^{\rm qssep}_k( \hat n_{i_1}\dots \hat n_{i_m})=\sum_{s=1}^m \Big({\cal L}^{\rm qssep}_k( \hat n_{i_s}) \prod_{\substack{r=1\\ r\neq s}}^m \hat n_{i_r}\Big) -\hat \Delta_k\sum_{1\leq s< q\leq m} \mathbf{1}_{\{i_s,i_q\}=\{k,k+1\}} \prod_{r\neq s,q} \hat n_{i_r}
\end{align}
with notation $\mathbf{1}_{\{i,j\}=\{k,l\}} :=\delta_{i,k}\delta_{j,l} + \delta_{j,k}\delta_{i,l}$. Assuming MFT scaling,
\be\label{eq:MFT-scaling}
\avg{\hat n_{i_1}\dots \hat n_{i_m}}^c ={\cal O}(N^{1-m})
\ee
one has that 
\be\label{eq:MFT-factorization}
\avg{ \hat P_k \hat n_{i_1}\dots \hat n_{i_m}} = \avg{ \hat P_k}\avg{ \hat n_{i_1}\dots \hat n_{i_m}}+{\cal O}(|S_N|/N),
\ee
if $\{i_1,\dots,i_m\}\cap(k+S_N)=\emptyset$. The proof of Eq.~\eqref{eq:MFT-factorization} easily follows. By writing $\hat P_k=\sqrt{D_0}(1+\lambda \Pt_k)$ and discarding the constant part, one has
\be
\kappa(\Pt_k,\hat n_{i_1},\dots ,\hat n_{i_m})=\sum_{r=1}^{|S_N|} \sum_{\substack{Q\subset S_N\\ r=|Q|}} \frac{b_r}{|S_N|^r}\Cum( \prod_{q\in Q} \hat n_{q+k},\hat n_{i_1},\dots,\hat n_{i_m})
\ee
Here, the sum over $Q\subset S_N$ is made of $\binom{|S_N| }{r} \leq |S_N|^r/r!$ terms, and $\Cum( \prod_{q\in Q} \hat n_{q+k},\hat n_{i_1},\dots,\hat n_{i_m})$ involves at least one density-density correlation pair, which scales as ${\cal O}(1/N)$ by assumption. The sum over $r$ then involves $|S_N|$ terms, so the connected part is at most of order ${\cal O}(|S_N|/N)$. 
\smsubsection{Density profile and diffusivity}
Specifying Eq.~\eqref{eq:single-rep-hydro-sm} to a single density operator, $\hat O_m\equiv \hat n_j$, one finds
\be
\frac{d}{dt}\avg{\hat n_j}=-\avg{\hat{J}_j -\hat J_{j-1}} 
\ee
where $\hat J_j:=-\hat D_j (\hat n_{j+1}-\hat n_j)$ is the current, and  $\hat D_j:= \hat P_j^2$. Since $\{0,1\}\not\in S_N$, we can use Eq.~\eqref{eq:MFT-factorization}, together with the gradient structure of the current, and obtain the closed equation
\be
\frac{d}{dt}\avg{\hat n_j}\simeq \avg{\hat D_j}\avg{\hat n_{j+1}-\hat n_j} - \avg{\hat D_{j-1}}\avg{\hat n_{j}-\hat n_{j-1}}.
\ee
In the limit $N\to\infty$ at fixed $x=j/N$ and $\tau=t/N^2$, defining
\be
\bar{n}(x;\tau):=\lim_{N\to\infty} \avg{\hat n_{j} }_{t}\Big\vert_{\substack{x=j/N;\\ \tau=t/N^2}}, \quad D(x;\tau):= \lim_{N\to\infty} \avg{\hat D_{j}}_{t}\Big\vert_{\substack{x=j/N;\\ \tau=t/N^2}}
\ee
one obtains
\be
\de_\tau \bar{n}(x;\tau)=\de_x \left( D(x;\tau)\de_x \bar{n}(x;\tau)\right).
\ee
It is possible to explicitly compute the diffusion constant $D(x;\tau)$. Starting from the microscopic definition of $\hat P_k$ in Eqs.~\eqref{eq:P-dressing} and \eqref{eq:P-subext} of the main text, one finds
\begin{align}
\avg{\hat D_j}=&D_0\lambda^2\sum_{r,s=0}^{|S_N|} \sum_{\substack{Q,K\subset S_N \\ |Q|=r, |K|=s} } \frac{b_r b_s }{|S_N|^{r+s}} \avg{\prod_{q\in Q\cup K} \hat n_{j+q}},\nn
\end{align}
with the convention that $b_0=1/\lambda$. Using that $\avg {\hat n_{i_1}\dots \hat n_{i_m}}= \prod_{r=1}^m \avg{\hat n_{i_r}} +{\cal O}(1/N)$,
\begin{align}
\avg{\hat D_j}=&D_0\lambda^2\sum_{r,s=0}^{|S_N|} \sum_{\substack{Q,K\subset S_N \\ |Q|=r, |K|=s} } \frac{b_r b_s }{|S_N|^{r+s}}\prod_{q\in Q\cup K}  \avg{\hat n_{j+q}}\left(1+{\cal O}(N^{-1})\right)\nn
=&D_0\lambda^2\sum_{r,s=0}^{|S_N|} \sum_{\substack{Q,K\subset S_N \\ |Q|=r, |K|=s} } \frac{b_r b_s }{|S_N|^{r+s}} \avg{\hat n_{j}}^{|Q\cup K|} \left(1+{\cal O}(|S_N|/N)\right).
\end{align}
In the limit $N\to\infty$, the diffusivity is therefore (cf.~Eq.~\eqref{eq:Diff-const} of the main text)
\be
D(x; \tau)=\sum_{m=0}^\infty B_m \ \bar{n}(x;\tau)^m , \qquad B_m:=\lim_{N\to\infty}\sum_{\substack{Q,K\subset S_N \\ |Q\cup K|=m}} \frac{D_0\lambda^2 b_{|Q|} b_{|K|}}{|S_N|^{|Q|+|K|}}.
\ee
with the convention that $b_0=1/\lambda$. Note that $D(x; \tau)\equiv D[\bar{n}(x;\tau)]$ is a generic function of the averaged density.\\

For the short-range example in Eq.~\eqref{eq:short-range-dressing} of the main text, 
\[
\Pt_k=\hat n_{k-1}(1-\hat n_{k+2})+\hat n_{k+2}(1-\hat n_{k-1}), 
\]
which corresponds to $S_N=\{-1,2\}$, $|S_N|=2$, and $b_1=2$, $b_2=-8$ in the general formula. The diffusivity coefficients then reads
\begin{align}
&B_0=D_0;\nn
&B_1=2D_0\lambda^2\left(\frac{b_0b_1}{2}+\frac{b_1b_0}{2}+\frac{b_1^2}{4}\right)=2D_0(2\lambda+\lambda^2);\nn
&B_2=D_0\lambda^2\left(\frac{b_0b_2}{2}+\frac{b_1^2}{2}+\frac{b_1b_2}{2}+\frac{b_2^2}{16}\right)=-2D_0(2\lambda+\lambda^2);\nn
&B_{m>2}=0.
\end{align}
Hence, for the short-range example,
\be
D[\bar n]=B_0+B_1\bar n+B_2\bar n^2=D_0\left[1+2(2\lambda+\lambda^2)\bar n(1-\bar n)\right].
\ee
Equivalently, one can obtain this result from a direct calculation. Using Eq.~\eqref{eq:short-range-dressing} of the main text and that $\Pt_j^2=\Pt_j$, one obtains
\be
\avg{\Pt_j}_t=\sigma_0[\avg{\hat n_j}_t] +{\cal O}(1/N),
\ee
given the definition of the SEP mobility $\sigma_0[n]:=2n(1-n)$. Therefore,
\be
\avg{\hat D_j}_t=D_0\avg{1+\big(2\lambda+\lambda^2\big) \Pt_j}_t=D_0\left(1+(2\lambda+\lambda^2)\sigma_0[\avg{\hat n_i}_t]\right) +{\cal O}(1/N).
\ee
\smsubsection{Connected density-density correlation}
We now discuss the connected density-density correlation. To begin with, from Eq.~\eqref{eq:evaluate-Lqseep-density} one finds,
\begin{align}
\frac{d}{dt}\avg{ \hat n_i \hat n_j }_t
=&-\avg{ \big(\hat J_i-\hat J_{i-1}\big) \hat n_j} -\avg{\big(\hat J_j -\hat J_{j-1}\big)\hat n_i  }_t \nn
&-\big(\delta_{i-1,j}-\delta_{ij}\big) \avg{\hat D_{i-1}(\hat n_i -\hat n_{i-1})^2}_t - \big(\delta_{i,j-1}-\delta_{ij}\big) \avg{\hat D_{i}(\hat n_{i+1} - \hat n_{i})^2}_t.
\end{align}
Next, we take the connected part. It is convenient to use the cumulant notation $\kappa(A,B)_t=\avg{A B}_t-\avg{A}_t\avg{B}_t$, so that the grouping of variables is unambiguous. Using the cumulant identity,
\be\label{cum-id-tmp}
\Cum(AB,C)=\Cum(A,B,C)+\Cum(A)\Cum(B,C)+\Cum(B)\Cum(A,C)
\ee
the previous expression becomes
\begin{align}\label{eq:dens-dens-tmp}
\frac{d}{dt}\avg{ \hat n_i \hat n_j }^c_t=&\Cum(\hat D_i, \hat n_{i+1}-\hat n_i, \hat n_j)_t-\Cum(\hat D_{i-1}, \hat n_{i}-\hat n_{i-1}, \hat n_j)_t\nn
&+\avg{\hat n_{i+1}-\hat n_i}_t \ \Cum(\hat D_i,\hat n_j)_t - \avg{\hat n_{i}-\hat n_{i-1}}_t\ \Cum(\hat D_{i-1},\hat n_j)_t\nn
&+\avg{\hat D_i}_t\ \Cum\left(\hat n_{i+1}-\hat n_{i},\hat n_j\right)_t -\avg{\hat D_{i-1}}_t\ \Cum\left(\hat n_{i}-\hat n_{i-1},\hat n_j\right)_t +(i\leftrightarrow j)\nn
&-\big(\delta_{i-1,j}-\delta_{ij}\big) \avg{\hat D_{i-1}(\hat n_i -\hat n_{i-1})^2}_t - \big(\delta_{i,j-1}-\delta_{ij}\big) \avg{\hat D_{i}(\hat n_{i+1} - \hat n_{i})^2}_t.
\end{align}

\textbf{\textit{1)~Discrete contact terms.}}~We first look at the discrete contact terms, which are simpler to analyze. These terms are
\begin{align}
\text{discrete contact}=&-\big(\delta_{i-1,j}-\delta_{ij}\big) \avg{\hat D_{i-1}(\hat n_i -\hat n_{i-1})^2}_t - \big(\delta_{i,j-1}-\delta_{ij}\big) \avg{\hat D_{i}(\hat n_{i+1} - \hat n_{i})^2}_t\nn
&= \big(\delta_{ij}-\delta_{i,j-1}\big)\hat Q_{i}-\big(\delta_{i,j+1}-\delta_{ij}\big)\hat Q_{i-1} = \nabla_i^{(-)}\nabla_j^{(-)} \hat C_{ij}, 
\end{align}
where $\nabla_i^{(-)}$ denote the discrete derivative, $\nabla_i^{(-)}f_i=f_{i}-f_{i-1}$ and  $\hat C_{ij}:= \delta_{i,j} \hat Q_i$ with $\hat Q_i:=\avg{\hat D_{i}(\hat n_{i+1} - \hat n_{i})^2}_t$. Using the factorization~\eqref{eq:MFT-factorization}, we find 
\begin{align}
\hat Q_i
:=D[\avg{\hat n_i}_t]\sigma_0[\avg{\hat n_i}_t]+{\cal O}(|S_N|/N).
\end{align}
In the scaling limit, using 
\be
\lim_{N\to\infty} N \delta_{ij}\Big\vert_{\substack{x=i/N\\ y=j/N}}=\delta(x-y),
\ee
one obtains
\be\label{eq:discr-cont-dens-dens}
\text{discrete contact}=N^{-3}\de_x\de_y\Big(D[\bar{n}(x; \tau)]\sigma_0[\bar{n}(x; \tau)]\delta(x-y)\Big).
\ee
\indent
\textbf{\textit{ 2)~Scaling limit.}}~The scaling $N^{-3}$ of the discrete contact terms also fixes the scaling of the connected density-density correlation. Since all the cumulants appearing in Eq.~\eqref{eq:dens-dens-tmp} are embedded in a double lattice-derivative, the reference scaling to leading order is set to $1/N$; this agrees also with the expected MFT scaling of the correlation. Therefore, we denote the scaling function of connected density-density correlations as
\be
C_2(x,y; \tau ):=\lim_{N\to \infty} N \avg{\hat n_{i} \hat n_{j}}^c_{t}\Big\vert_{\substack{x=i/N\\ y=j/N\\ t=\tau N^2}}\equiv  \lim_{N\to \infty} N \Cum(\hat n_{i} , \hat n_{j})_{t}\Big\vert_{\substack{x=i/N\\ y=j/N\\ t=\tau N^2}}.
\ee
Note that, by rewriting
$\Cum(\hat n_i,\hat n_j)=\frac12\delta_{ij}\sigma_0[\avg{\hat n_i}] +\Cum_\text{off}(\hat n_i,\hat n_j)$, with off-diagonal part $\Cum_\text{off}(\hat n_i,\hat n_j):=(1-\delta_{ij})\Cum(\hat n_i,\hat n_j)$, one has
\begin{align}
\lim_{N\to\infty} N \ \frac12 \delta_{ij} \sigma_0[\avg{\hat n_i}] \Big\vert_{\substack{x=i/N\\ y=j/N\\ t=\tau N^2}}\ = \delta(x-y) \frac{1}{2}\sigma_0[\bar{n}(x;\tau)];\qquad 
\lim_{N\to\infty} N    \ \Cum_\text{off}(\hat n_i,\hat n_j)\Big\vert_{\substack{x=i/N\\ y=j/N\\ t=\tau N^2}} \ = C_{2;\text{reg}}(x,y;\tau);
\end{align}
and the scaling function of interest is written as a regular part plus a contact contribution,
\be
C_2(x,y;\tau)=C_{2;\text{reg}}(x,y;\tau)+ \delta(x-y) \frac{1}{2}\sigma_0[\bar{n}(x;\tau)].
\ee
We therefore distinguish the diagonal $i=j$ from the coinciding-point limit of distinct sites, $i\neq j$ with $|i-j|/N\to0$. The second contribution defines $C_{2,\mathrm{reg}}(x,x;\tau)$, while the first one gives the contact term proportional to $\delta(x-y)$.\\

\indent
\textbf{\textit{ 3)~Bulk contribution.}}~We now evaluate the bulk contributions in Eq.~\eqref{eq:dens-dens-tmp}. There are three such terms involving cumulants of density operators. To leading order in $1/N$, one can show that
\begin{align}\label{eq:cum-type0-dens-dens}
\avg{\hat D_i}_t\Cum\left(\hat n_{i+1}-\hat n_{i},\hat n_j\right)_t -\avg{\hat D_{i-1}}_t\Cum\left(\hat n_{i}-\hat n_{i-1},\hat n_j\right)_t +(i\leftrightarrow j) &\simeq N^{-3}\de_x \left({D}[\bar{n}(x;\tau)]\de_x C_2(x,y;\tau)\right) + (x\leftrightarrow y);\\[5pt]
\avg{\hat n_{i+1}-\hat n_i}_t \ \Cum(\hat D_i,\hat n_j)_t - \avg{\hat n_{i}-\hat n_{i-1}}_t\ \Cum(\hat D_{i-1},\hat n_j)_t +(i\leftrightarrow j) &\simeq  N^{-3} \de_x\left( \left(\de_x D[\bar{n}(x;\tau)]\right) C_2(x,y;\tau)\right) + (x\leftrightarrow y);\label{eq:cum-type1-dens-dens}\\[5pt]
\Cum(\hat D_i, \hat n_{i+1}-\hat n_i, \hat n_j)_t-\Cum(\hat D_{i-1}, \hat n_{i}-\hat n_{i-1}, \hat n_j)_t +(i\leftrightarrow j) &={\cal O}(N^{-4}).
\end{align}
These estimates follow from Eq.~\eqref{eq:MFT-factorization}. In particular, for far apart points, $|i-j|\gg |S_N|$, a direct application of \eqref{eq:MFT-factorization} gives
\[
\Cum(\hat D_i,\hat n_j)_t=D'[\avg{\hat n_i}_t]\ \Cum_\text{off}(\hat n_i,\hat n_j)_t \left(1+ {\cal O}(|S_N|/N)\right), \qquad |i-j|\gg |S_N|,
\]
together with the discrete gradients in $i$. Contact configurations, where $j-i\in S_N\cup\{0,\pm1\}$, require a separate treatment. For instance,
\[
\Cum(\hat D_i,\hat n_j)_t=\frac{\chi_{\{ j-i\}\in S_N}}{|S_N|} D'[\avg{\hat n_i}_t] \ \frac12\sigma_0[\avg{\hat n_j}_t]  \left(1 + {\cal O}(|S_N|/N)\right),  \qquad \{j-i\}\in S_N.
\]
Here, $\chi_{\textit{cond}}:=\{1, \;\text{if {\it cond}$=$true}; \; 0, \; \text{otherwise}\}$ is the indicator function of the support. In particular, $\lim_{N\to \infty} \frac{N}{|S_N|} \chi_{\{ j-i\}\in S_N}= \delta(x-y)$. These terms thus contribute to the diagonal regular and contact parts of Eqs.~\eqref{eq:cum-type0-dens-dens}--\eqref{eq:cum-type1-dens-dens}.\\

Combining the contact terms~\eqref{eq:discr-cont-dens-dens} with \eqref{eq:cum-type0-dens-dens}-\eqref{eq:cum-type1-dens-dens} in Eq.~\eqref{eq:dens-dens-tmp}, one obtains Eq.~\eqref{eq:C-hydro} of the main text. 
\smsection{Multi-replica coherences}\label{smsec:multi-replica}
We now specialize to the $n$-point coherence loop 
\be
\G_{i_1,i_2,\dots,i_n}(t):=\E[\prod_{r=1}^n \langle \hat X_{i_r,i_{r+1}}\rangle_t], \qquad (i_{n+1}\equiv i_1)
\ee
where, for simplicity, we take $i_1,i_2,\dots, i_n$ all distinct. We consider the scaling
\be
g_n(x_1,\dots, x_n;\tau):=\lim_{N\to \infty} N^{n-1} \G_{i_1,i_2,\dots,i_n}(t)\ \Big\vert_{\substack{x_k=i_k/N \\ \tau=t/N^2}}.
\ee
As shown below, this scaling of the coherence loop is compatible with the structure of the contact terms, and is ultimately justified by the MFT scaling of density operators in Eq.~\eqref{eq:MFT-scaling}. Following the notation introduced in the \hyperlink{EM}{End Matter}, the lattice equation of motion is
\begin{align}
\frac{d}{dt} \G_{i_1,i_2,\dots,i_n}(t)=&dt \sum_{r=1}^n\sum_{k=0}^{N-1}\E\left[\left\langle \hat P_k {\cal L}_k^\text{qssep}(\hat X_{i_r, i_{r+1}})\hat P_k- {\cal D}_k(\hat X_{i_r,i_{r+1}}) \right\rangle_t \prod_{s\neq r} \langle \hat X_{i_s,i_{s+1}}\rangle_t\right]\nn
&-\sum_{1\leq r<q\leq n} \sum_{k=0}^{N-1}\E\left[\left\langle d{\cal H}_k(\hat X_{i_r,i_{r+1}})\right\rangle_t\left\langle d{\cal H}_k(\hat X_{i_q,i_{q+1}})\right\rangle_t \prod_{s\neq r,q} \langle \hat X_{i_s,i_{s+1}}\rangle_t\right].
\end{align}
It is useful to organize the calculation as follows. We define the single-replica contributions
\begin{align}
\circled{a}_{ij;k}:=&\langle \hat P_k {\cal L}_k^\text{qssep}(\hat X_{ij})\hat P_k\rangle_t;\\[5pt]
\circled{b}_{ij;k}:=&\langle  {\cal D}_k(\hat X_{ij})\rangle_t,
\end{align}
and the two-replica It\=o contractions,
\begin{align}
\circled{c}_{ijlm;k}:=&\frac{1}{dt} \ \langle \hat P_k [d\hat{h}_k,\hat X_{ij}]\rangle_t\langle \hat P_k [d\hat{h}_k,\hat X_{lm}]\rangle_t;\\[5pt]
\circled{d}_{ijlm;k}:=&\frac{1}{dt} \ \langle \hat P_k [d\hat{h}_k,\hat X_{ij}]\rangle_t\langle [\hat P_k,\hat X_{lm}]d\hat{h}_k \rangle_t + (ij \leftrightarrow lm);\\[5pt]
\circled{e}_{ijlm;k}:=&\frac{1}{dt} \ \langle [\hat P_k ,\hat X_{ij}]d\hat{h}_k\rangle_t\langle [\hat P_k ,\hat X_{lm}]d\hat{h}_k\rangle_t.
\end{align}
Since we look at diffusive scaling regime, the target scaling for each of these term is ${\cal O}(N^{-n-1})$.\\

We introduce the notation
\be
\de_{\hat n_a}\hat P_k:=\hat P_k\big|_{\hat n_a=1}-\hat P_k\big|_{\hat n_a=0},
\ee
together with
\begin{align}
\Pnot{a}_k:=&\hat P_k\big|_{\hat n_a=0}=\hat P_k-\hat n_a\,\de_{\hat n_a}\hat P_k;\\[6pt]
\Pyes{a}_k:=&\hat P_k\big|_{\hat n_a=1}=\hat P_k+(1-\hat n_a)\de_{\hat n_a}\hat P_k= \Pnot{a}_k + \de_{\hat n_a}\hat P_k.
\end{align}
Thus $\Pnot{a}_k$ denotes the polynomial $\hat P_k$ evaluated with site $a$ empty, while $\Pyes{a}_k$ denotes the same polynomial evaluated with site $a$ occupied. With this notation, one has
\be
[\hat P_k,\hat X_{ij}]=\left( \de_{\hat n_i} \Pnot{j}_k-\de_{\hat n_j}\Pnot{i}_k\right)\hat X_{ij}.
\ee
From the definitions in Eqs.~\eqref{eq:P-dressing} and \eqref{eq:P-subext} of the main text, one has
\be
\de_{\hat n_i}\hat P_k=\left( \lambda\sqrt{D_0}\;\frac{\chi_{\{i-k\}\in S_N}}{|S_N|}\right)\Bigg(\sum_{r\geq 1}\; \sum_{\substack{Q\subset S_N\setminus\{i-k\} \\ |Q|=r-1}}\frac{b_r}{|S_N|^{r-1}}\prod_{q\in Q} \hat n_{k+q}\Bigg)=\chi_{\{i-k\}\in S_N} \times {\cal O}(\lambda/|S_N|).
\ee
Here, $\chi_{\{i-k\}\in S_N}$ is the indicator function on the support. Since $\lim_{N\to \infty} \frac{N}{|S_N|} \chi_{\{ j-i\}\in S_N}= \delta(x-y)$, terms involving $\de_{\hat n}\hat P$ are supported on a mesoscopic window which collapses to coinciding mesoscopic points in the large-scale limit.
\smsubsection{Multi-replica factorization}
In the derivation below, we shall use the factorization hypothesis discussed in Eq.~\eqref{eq:factorization} of the main text and in Sec.~\ref{sec:factorization-hyp} of the \hyperlink{EM}{End Matter}. Before applying the factorization ansatz, we first reduce any overlap between the support of the density polynomial and that of the coherence operator. For instance,
\be
\hat F(\hat n) \hat X_{i_q,i_{q+1}}=\hat F^{(i_q,\neg i_{q+1})}(\hat n)\hat X_{i_q,i_{q+1}},
\ee
with $\hat F^\text{(red)}\equiv\hat F^{(i_q,\neg i_{q+1})}$. Then, away from contact configurations, the multi-replica factorization ansatz can be stated as
\be\label{eq:SM-factorization}
\E\Big[\langle \hat F^\text{(red)}(\hat n) \hat X_{i_q,i_{q+1}}\rangle_t \prod_{\substack{r=1 \\ r\neq q}}^n\langle\hat X_{i_r,i_{r+1}}\rangle_t\Big]=\avg{\hat F^\text{(red)}(\hat n)}_t\; \E\Big[\prod_{r=1}^n \langle\hat X_{i_r,i_{r+1}}\rangle_t\Big] +{\cal O}(N^{-n-\nu_F})
\ee
with $i_{n+1}\equiv i_1$, and $\avg{\hat F^\text{(red)}}={\cal O}(N^{-\nu_F})$. For multiple insertions of density polynomials $\hat F_a(\hat n)$, the error is ${\cal O}(N^{-n-\sum_a \nu_{F_a}})$. As discussed in the \hyperlink{EM}{End Matter}, the error is dynamically stable, and its scaling exponent is fixed by the leading contact terms.

As a simple explicit check, we consider a two-replica coherence loop with a density insertion on the first replica, cf.~Eq.~\eqref{eq:factorization} of the main text. We thus consider the factorization error,
\be
Y_{k;ij}(t):=\E[\langle \hat n_k \hat X_{ij}\rangle_t\langle\hat X_{ji}\rangle_t]-\avg{\hat n_k}_t \ \G_{ij}(t),
\ee
with $Y_{k;ij}={\cal O}(N^{-\gamma})$. We need to show that: {\it i)}~the error is dynamically stable; {\it ii)}~the scaling exponent is $\gamma=2$. First, we note that, away from contact configurations, the martingale of $\hat n_k$, which is supported on the links $k$, $k-1$,  has vanishing It\=o contractions with the martingales of $\hat X_{ij}$ and $\hat X_{ji}$. The equation of motion for $Y_{k;ij}$ therefore has the same diffusive structure as the equations for the density and the coherence loop. As a result,  the scaling $Y_{k;ij}={\cal O}(N^{-\gamma})$ is preserved on diffusive time scales.\\

The scaling exponent $\gamma$ is fixed by the contact terms. With simple algebra, one obtains
\begin{align}
\frac{d}{dt}Y_{k;ij}\Big\vert_\text{$k\simeq j$}\simeq&D_0 \Big[ \delta_{k-1,j}
\Big(\avg{n_{j+1}}\ \G_{ij}+\avg{n_j}\ \G_{i,j+1}\Big)
+\delta_{k+1,j}\Big(\avg{n_{j-1}}\ \G_{ij}+\avg{n_j}\ \G_{i,j-1}\Big)\nn
&\qquad -\delta_{k,j}\Big(\avg{n_{j-1}+n_{j+1}}\ \G_{ij}+\avg{n_j}\left[\G_{i,j+1}+\G_{i,j-1}\right]\Big)\Big]={\cal O}(N^{-4}),
\end{align}
and analogously for $k$ approaching site $i$. This yields $\gamma=2$, in agreement with Eq.~\eqref{eq:factorization}.
\smsubsection{QSSEP-like drift and It\=o contractions}
By QSSEP-like terms we refer to the terms $\circled{a}_{ij;k}$ and $\circled{c}_{ijlm;k}$ defined above, which reduce to the standard QSSEP dynamics upon setting $\lambda=0$. We start from the Lindbladian part,
\begin{align}
\circled{a}_{ij;k}=&-\frac12\left(\delta_{k,i}+\delta_{k,i-1}+\delta_{k,j}+\delta_{k,j-1}\right) \langle\hat P_k \hat X_{ij} \hat P_k\rangle_t\nn
=& -\frac12\left\langle\left(\delta_{k,i} \hat D^{[j]}_i+\delta_{k,i-1} \hat D^{[j]}_{i-1}+\delta_{k,j} \hat D^{[i]}_j+\delta_{k,j-1}\hat D^{[i]}_{j-1}\right) \hat X_{ij}\right\rangle_t,
\end{align}
with $\hat D_k^{[a]}:=\Pnot{a}_k\Pyes{a}_k$, and $\hat D_k^{[a]}\equiv\hat D_k=\hat P_k^2$ whenever $\{a-k\}\notin S_N$. The It\=o contraction is
\begin{align}
\circled{c}_{ijlm;k}=&\delta_{k,i}\delta_{l,i+1}\langle \Pnot{j}_i \hat X_{i+1,j}\rangle_t\langle \Pnot{m}_i \hat X_{im}\rangle_t-\delta_{k,i}\delta_{i,m}\langle \Pnot{j}_i \hat X_{i+1,j}\rangle_t\langle \Pyes{l}_i \hat X_{l,i+1}\rangle_t\nn
&-\delta_{k,j-1}\delta_{j,l}\langle\Pyes{i}_{j-1}\hat X_{i,j-1}\rangle_t\langle \Pnot{m}_{j-1}\hat X_{j-1,m}\rangle_t+\delta_{k,j-1}\delta_{j-1,m}\langle\Pyes{i}_{j-1}\hat X_{i,j-1}\rangle_t\langle \Pyes{l}_{j-1}\hat X_{l,j}\rangle_t\nn
&+(ij\leftrightarrow lm).
\end{align}
\smsubsection{Drift correction}
Next, we evaluate the drift correction, $\circled{b}_{ij;k}$. A straightforward calculation gives
\begin{align}\label{eq:b-term}
\circled{b}_{ij;k}=&-\frac12\Big\langle\Big(\delta_{k,j-1} \left(\de_{\hat n_i} \hat P_{j-1}\right) \hat P_{j-1} (1-2\hat n_{j-1}) + \delta_{k,j}\left(\de_{\hat n_i}\hat P_j\right)\hat P_j (1-2\hat n_{j+1})
 \nn
&+\delta_{k,i-1}\left(\de_{\hat n_j} \hat P_{i-1}\right) \hat P_{i-1} (1-2\hat n_{i-1}) + \delta_{k,i} \left(\de_{\hat n_j}\hat P_i\right)\hat P_i (1-2\hat n_{i+1})\Big)\hat X_{ij}\Big\rangle_t\nn
&-\Big\langle(\de_{\hat n_i} \Pnot{j}_k)(\de_{\hat n_j} \Pnot{i}_k) \hat\Delta_k \hat X_{ij}\Big\rangle_t+\frac12\Big\langle\big((\de_{\hat n_i} \Pnot{j}_k)^2+(\de_{\hat n_j} \Pnot{i}_k)^2\big) \hat\Delta_k^{(i,\neg j)}\hat X_{ij}\Big\rangle_t,
\end{align}
where we recall $\hat \Delta_k=(\hat n_{k+1} - \hat n_k)^2$, and
\be
 \hat\Delta_k^{(i,\neg j)}=\begin{cases}
 1-\hat n_{i+1}, & k=i;\\
  1-\hat n_{i-1}, & k=i-1;\\
   \hat n_{j+1}, & k=j;\\
    \hat n_{j-1}, & k=j-1;\\
    \hat\Delta_k,& \text{otherwise}.
 \end{cases}
\ee
For $|i-j|\gg |S_N|$, we note that only the last term in Eq.~\eqref{eq:b-term} survives. Therefore,
\be
\circled{b}_{ij;k}=\frac12\Big\langle\big((\de_{\hat n_i} \Pnot{j}_k)^2+(\de_{\hat n_j} \Pnot{i}_k)^2\big) \hat\Delta_k^{(i,\neg j)}\hat X_{ij}\Big\rangle_t +\text{meso-hydro contact terms},
\ee
where meso-hydro contacts denote the terms that are non-vanishing when $|i-j|/N\to0$ in the large-scale limit. Moreover, these contact configurations
\begin{align}
\circled{b}_{ij;k}\Big\vert_\text{meso-hydro contact}=&-\frac12\Big\langle\Big(\delta_{k,j-1} \left(\de_{\hat n_i} \hat P_{j-1}\right) \hat P_{j-1} (1-2\hat n_{j-1}) + \delta_{k,j}\left(\de_{\hat n_i}\hat P_j\right)\hat P_j (1-2\hat n_{j+1})
 \nn
&+\delta_{k,i-1}\left(\de_{\hat n_j} \hat P_{i-1}\right) \hat P_{i-1} (1-2\hat n_{i-1}) + \delta_{k,i} \left(\de_{\hat n_j}\hat P_i\right)\hat P_i (1-2\hat n_{i+1})\Big)\hat X_{ij}\Big\rangle_t\nn
&+\text{quadratic in $\de_{\hat n}\hat P$},
\end{align}
and the meso-hydro contact terms quadratic in $\de_{\hat n}\hat P$ can be neglected since they scale as $\sim N^{-n-2}$.
\smsubsection{Additional It\=o contractions}
Finally, we move to the additional It\=o contractions. After simple algebra, these become
\begin{align}
\circled{d}_{ijlm;k}=& \delta_{k,i}  \langle \Pnot{j}_i \hat X_{i+1,j}\rangle_t\langle  \big(\de_{\hat n_l} \Pnot{m}_i -\de_{\hat n_m} \Pnot{l}_i\big) \wp[\hat X_{lm} \hat X_{i,i+1}]\rangle_t\nn
&-\delta_{k,j-1}\langle \Pyes{i}_{j-1} \hat X_{i,j-1} \rangle_t\langle  \big(\de_{\hat n_l} \Pnot{m}_{j-1} -\de_{\hat n_m} \Pnot{l}_{j-1}\big) \wp[\hat X_{lm} \hat X_{j-1,j}]\rangle_t\nn
&+  \delta_{k,i-1}\langle \Pnot{j}_{i-1} \hat X_{i-1,j}\rangle_t\langle  \big(\de_{\hat n_l} \Pnot{m}_{i-1} -\de_{\hat n_m} \Pnot{l}_{i-1}\big) \wp[\hat X_{lm} \hat X_{i,i-1}]\rangle_t\nn
&- \delta_{k,j} \langle \Pyes{i}_j\hat X_{i,j+1} \rangle_t\langle  \big(\de_{\hat n_l} \Pnot{m}_j -\de_{\hat n_m} \Pnot{l}_j\big) \wp[\hat X_{lm} \hat X_{j+1,j}]\rangle_t +(ij\leftrightarrow lm);
\end{align}
and
\begin{align}
\circled{e}_{ijlm;k}=& \left\langle \left(\de_{\hat n_i} \Pnot{j}_k -\de_{\hat n_j} \Pnot{i}_k\right) \wp[\hat X_{ij} \hat\ell_k] \right\rangle_t  \left\langle \left(\de_{\hat n_l} \Pnot{m}_k -\de_{\hat n_m} \Pnot{l}_k\right) \wp[\hat X_{lm}\hat\ell^\dagger_k]\right\rangle_t\nn
& +\left\langle \left(\de_{\hat n_i} \Pnot{j}_k -\de_{\hat n_j} \Pnot{i}_k\right) \wp[\hat X_{ij}\hat\ell^\dagger_k]\right\rangle_t  \left\langle \left(\de_{\hat n_l} \Pnot{m}_k -\de_{\hat n_m} \Pnot{l}_k\right) \wp[\hat X_{lm}\hat\ell_k]  \right\rangle_t.
\end{align}
Here the symbol $\wp[\dots]$ denotes the reduction of products of coherences whenever contact points are present. In particular, one has
\be\label{eq:contact-red}
\wp[\hat X_{ij}\hat X_{lm}]=\delta_{jl}(1-\delta_{im})(1-\hat n_j)\hat X_{im}-\delta_{im}(1-\delta_{jl})\hat n_i\hat X_{lj}+\delta_{jl}\delta_{im}\hat n_i(1-\hat n_j)+(1-\delta_{jl})(1-\delta_{im})\hat X_{ij}\hat X_{lm}.
\ee

We note that the terms $\circled{d}_{ijlm;k}$ and $\circled{e}_{ijlm;k}$ may generate products of coherences on each replica, thus potentially breaking the closure of the equation of motion for the coherence loop, even after using the factorization in Eq.~\eqref{eq:SM-factorization}. However, by setting $\lambda=c_\lambda\lambda_*$ and taking the limit $c_\lambda\to0$, the scaling of these terms can be inferred from the Wick counting at the QSSEP point. In particular, for distinct points and away from contact reductions, one finds for instance
\be
\frac{\E[\langle \hat X_{aj} \rangle_t \langle \hat X_{ia} \hat X_{lm}\rangle_t \dots]}{\E[\langle \hat X_{ij}\rangle_t \langle \hat X_{lm}\rangle_t \dots]}={\cal O}(1/N), \qquad 
\frac{\E[\langle \hat X_{ij} \hat X_{ab}\rangle_t \langle \hat X_{lm} \hat X_{ba}\rangle_t \dots]}{\E[\langle \hat X_{ij}\rangle_t \langle \hat X_{lm}\rangle_t \dots]}={\cal O}(1/N).
\ee
Contact reductions~\eqref{eq:contact-red}, in which a product of two coherences reduces to a single coherence dressed by densities, are not included in this counting. By scaling arguments, one then finds, away from contact reductions,
\be
\sum_k \circled{d}_{ijlm;k}={\cal O}\left(\frac{\lambda}{\lambda_*} N^{-n-2+\alpha/2}\right),
\ee
i.e. this contribution is subleading with respect to the target scaling $\sim N^{-n-1}$ when $\lambda\sim \lambda_*$. Contact reductions \eqref{eq:contact-red} may instead give contributions of order $\sum_k \circled{d}_{ijlm;k}\sim N^{-n-1+\alpha/2}$, which are potentially dangerous and must be analyzed separately.\\

By similar scaling arguments, terms quadratic in $\de_{\hat n}\hat P$ are always subleading. Indeed, one finds
\be
\sum_k \circled{e}_{ijlm;k}
=
{\cal O}\left(\frac{\lambda^2}{\lambda_*^2} N^{-n-2}\right).
\ee
In the following, we therefore neglect the contribution of $\circled{e}$ to the equation of motion for the coherence loop.\\

The expressions $\circled{a}$--$\circled{e}$ above have general validity. In the following section, we specify them to the two-replica case, while in Sec.~\ref{sec:sm-generalization} we discuss the generalization of the two-replica calculation to arbitrary coherence loops.
\smsection{Two-replica coherence loop}
We now specialize to the two-replica coherence loop $\G_{ij}(t):=\E[\langle\hat X_{ij}\rangle_t\langle \hat X_{ji}\rangle_t]$, with $i\neq j$ for simplicity. Using the notation introduced in the previous section, and neglecting the contribution of $\circled{e}$ as discussed above, the equation of motion reads
\be\label{eq:SM:2rep-loop-eom}
\frac{d}{dt}\G_{ij}(t)=\sum_k \E\left[\big(\circled{a}_{ij;k}-\circled{b}_{ij;k}\big)\langle \hat X_{ji}\rangle_t+\langle \hat X_{ij}\rangle_t\big(\circled{a}_{ji;k}-\circled{b}_{ji;k}\big)-\circled{c}_{ijji;k}-\circled{d}_{ijji;k}\right].
\ee
We first evaluate the QSSEP-like contribution, namely the combination of $\circled{a}$ and $\circled{c}$. Using Eq.~\eqref{eq:SM-factorization}, we obtain
\begin{align}\label{eq:SM-term1}
\rcircled{1}=&\sum_k \E[\circled{a}_{ij;k}\langle \hat X_{ji}\rangle_t+ \langle \hat X_{ij}\rangle_t \ \circled{a}_{ji;k} - \circled{c}_{ijji;k}]\nn
\overset{\eqref{eq:SM-factorization}}{\simeq}&
-\avg{ \hat D_i^{[j]}+ \hat D_{i-1}^{[j]} } \G_{ij}+  \avg{\Pyes{j}_i}\avg{\Pnot{j}_i} \G_{i+1,j}+ \avg{\Pyes{j}_{i-1}}\avg{\Pnot{j}_{i-1}} \G_{i-1,j}+(i\leftrightarrow j)\nn
&-2\delta_{i+1,j}\avg{\hat P_i}^2\E[\langle\hat n_{i+1}\rangle\langle\hat n_i\rangle] -2 \delta_{i-1,j}\avg{\hat P_{i-1}}^2 \E[\langle\hat n_{i}\rangle\langle\hat n_{i-1}\rangle]\nn
=& D_0\ \Delta_i \G_{ij}-2\big(\avg{\hat D_i}-\avg{\hat P_i}^2\big) \G_{ij}+(i\leftrightarrow j) +\delta_{i+1,j}\avg{\hat P_i}^2\big(\G_{ii}+\G_{i+1,i+1}-2\E[\langle\hat n_{i+1}\rangle\langle\hat n_i\rangle]\big) \nn
&+ \delta_{i-1,j}\avg{\hat P_{i-1}}^2\big(\G_{ii}+\G_{i-1,i-1}-2 \E[\langle\hat n_{i}\rangle\langle\hat n_{i-1}\rangle]\big) +{\cal O}(N^{-4+\alpha}),
\end{align}
where $\Delta_i f_i= f_{i+1}+f_{i-1}-2f_i$ is the discrete Laplacian. The last two terms in Eq.~\eqref{eq:SM-term1} give the usual QSSEP contact source, see \cite{Hruza2023}. Indeed, using the factorization of density operators, for instance $\G_{ii}\simeq \avg{\hat n_i}^2$, one has
\begin{align}
&\avg{ \hat P_i}^2\delta_{i+1,j}\big(\G_{ii} + \G_{i+1,i+1} -2 \E[\langle\hat n_{i}\rangle\langle \hat n_{i+1}\rangle]\big)+\avg{ \hat P_{i-1}}^2\delta_{i-1,j}\big(\G_{ii}+\G_{i-1,i-1} -2 \E[\langle\hat n_{i}\rangle\langle \hat n_{i-1}\rangle]\big)\nn
\simeq&\avg{ \hat P_i}^2\delta_{i+1,j}\big(\avg{\hat n_{i+1}}-\avg{\hat n_i}\big)^2
+\avg{ \hat P_j}^2\delta_{i-1,j}\big(\avg{\hat n_i}-\avg{\hat n_{i-1}}\big)^2\nn
=& 2N^{-3} D_0 \delta(x-y) \left(\de_x\bar{n}(x;\tau)\right)^2 \left(1+{\cal O}(\lambda)\right)=N^{-3}\ {\cal S}^{\rm qssep}_2(x,y;\tau).
\end{align}
Thus, the combination of $\circled{a}$ and $\circled{c}$ in Eq.~\eqref{eq:SM-term1} generates the Laplacian and the QSSEP contact source. In addition, it produces the first contribution in Eq.~\eqref{eq:mass-counting} of the main text, namely
\be
a^{(1)}_{i,t}:=2N^2\big(\avg{\hat D_i}_t-\avg{\hat P_i}_t^2\big). 
\ee
We now turn to the remaining terms in Eq.~\eqref{eq:SM:2rep-loop-eom}. They give
\begin{align}\label{eq:SM-term2}
\rcircled{2}=& - \sum_k \E[ \circled{b}_{ij;k}\langle\hat X_{ji}\rangle +\langle\hat X_{ij}\rangle\circled{b}_{ji;k} + \circled{d}_{ijji;k}] \overset{\eqref{eq:SM-factorization}}{\simeq} - \sum_k \avg{(\de_{\hat n_i} \hat P_k)^2  \hat\Delta_k} \  \G_{ij} +(i\leftrightarrow j) +{\cal O}(N^{-4+\alpha/2}),
\end{align}
where we recognize the second additional contribution in Eq.~\eqref{eq:mass-counting} of the main text,
\be
a^{(2)}_{i,t}:=N^2\sum_{k=0}^{N-1} \avg{(\de_{\hat n_i} \hat P_k)^2  \hat\Delta_k}_t.
\ee
Interestingly, the meso-hydro contact terms linear in $\de_{\hat n}\hat P$ coming from the $\circled{b}$ and $\circled{d}$ contributions cancel to leading order in $1/N$, leaving ${\cal S}_2^\text{qseep}(x,y)$ as the only contact source of the equation.\\

From Eqs.~\eqref{eq:SM-term1} and \eqref{eq:SM-term2}, and using the definition in Eq.~\eqref{eq:mass} of the main text, one recovers Eq.~\eqref{eq:g2-massive} of the main text for the evolution of the two-replica coherence loop.
\smsubsection{Microscopic expression of the mass term}
\label{sec:sm-mass}
In this section, we compute explicitly the mass term in Eq.~\eqref{eq:mass} of the main text in terms of the microscopic parameters entering the interaction dressing \eqref{eq:P-dressing}-\eqref{eq:P-subext} of the main text. We separate the two contributions introduced in Eq.~\eqref{eq:mass-counting}.\\

Recall,
\[
\hat P_i=\sqrt{D_0}\big(1+\lambda \Pt_i\big), \quad\text{with}\quad \Pt_i=\sum_{r\geq 1} \sum_{\substack{Q\subset S_N\\ |Q|=r}} \frac{b_r}{|S_N|^r} \prod_{q\in Q} \hat n_{i+q}.
\]
From these definitions, one obtains for the first term
\begin{align}
m^{2}_{(1)}(x;\tau):= \lim_{N\to\infty} D_0^{-1} a^{(1)}_{i,t}\Big\vert_{\substack{ x=i/N \\ \tau=t/N^2}}= 2 \lim_{N\to \infty} N^2 \lambda^2 \Cum(\Pt_i,\Pt_i)\Big\vert_{\substack{i=xN\\ t=\tau N^2}}.
\end{align}
The cumulant reads
\begin{align}
\Cum(\Pt_i,\Pt_i)&=\sum_{r,s\geq 1} \sum_{\substack{Q,K\subset S_N\\ |Q|=r,\ |K|=s}} \frac{b_r b_s}{|S_N|^{r+s}} \ \Cum\big(\prod_{q\in Q}\hat n_{i+q},\prod_{k\in K}\hat n_{i+k}\big).
\end{align}
Using the MFT factorization~\eqref{eq:MFT-factorization}, and to leading order in $1/N$, the cumulant becomes
\be
\Cum\big(\prod_{q\in Q}\hat n_{i+q},\prod_{k\in K}\hat n_{i+k}\big)=\avg{\hat n_i}^{|Q\cup K|}-\avg{\hat n_i}^{|Q|+|K|}+{\cal O}(|S_N|/N).
\ee
Therefore,
\begin{align}
\Cum(\Pt_i,\Pt_i)&=\sum_{r,s\geq 1} \sum_{\substack{Q,K\subset S_N\\ |Q|=r,\ |K|=s}} \frac{b_r b_s}{|S_N|^{r+s}}\left(\avg{\hat n_i}^{|Q\cup K|}-\avg{\hat n_i}^{r+s}\right)+{\cal O}(|S_N|/N).
\end{align}
We now group the pairs $Q,K\subset S_N$ according to their overlap $\ell:=|Q\cap K|$. For fixed $(r,s,\ell)$, one has $|Q\cup K|=r+s-\ell$, and therefore
\be
\avg{\hat n_i}^{|Q\cup K|}-\avg{\hat n_i}^{r+s} = \avg{\hat n_i}^{r+s-\ell} \left(1-\avg{\hat n_i}^{\ell}\right) = \frac12\sigma_0[\avg{\hat n_i}] \sum_{p=0}^{\ell-1} \avg{\hat n_i}^{r+s-\ell-1+p}.
\ee
Thus, a given overlap sector contributes to all powers $m=r+s-\ell-1,\dots,r+s-2$ of the density. It follows that,
\begin{align}
\Cum(\Pt_i,\Pt_i) &= \frac{1}{2|S_N|} \sigma_0[\avg{\hat n_i}]\Big(\sum_{m\geq0} A^{(1)}_m \avg{\hat n_i}^m \Big)+{\cal O}(|S_N|/N),
\end{align}
with
\be\label{eq:SM-Am1-coef}
A^{(1)}_m=\sum_{\substack{r,s\geq1\\ r+s\geq m+2}} \sum_{\ell=\max(1,r+s-m-1)}^{\min(r,s)} \frac{b_r b_s}{|S_N|^{r+s-1}}\, {\cal N}_{r,s,\ell}.
\ee
Here,
\be
{\cal N}_{r,s,\ell}=
\binom{|S_N|}{\ell}
\binom{|S_N|-\ell}{r-\ell}
\binom{|S_N|-r}{s-\ell}
\ee
counts the number of pairs of subsets $Q,K\subset S_N$ such that $|Q|=r$, $|K|=s$, and $|Q\cap K|=\ell$. Binomial coefficients with inadmissible entries are understood to vanish. In the scaling limit,
\be
m^{2}_{(1)}(x;\tau)
=
c_\lambda\,
\sigma_0[\bar n(x;\tau)]
\Big(
\sum_{m\geq0}A^{(1)}_m\,\bar n(x;\tau)^m
\Big),  \quad c_\lambda=\lambda^2/\lambda_*^2={\cal O}(1).
\ee
We now turn to the second contribution, defined as
\be
m^{2}_{(2)}(x;\tau):= \lim_{N\to\infty} D_0^{-1} a^{(2)}_{i,t}\Big\vert_{\substack{ x=i/N \\ \tau=t/N^2}}=\lim_{N\to\infty} N^2\lambda^2 \sum_{k:\{i-k\}\in S_N} \avg{(\de_{\hat n_i}\Pt_k)^2\hat\Delta_k} \Big\vert_{\substack{i=xN\\ t=\tau N^2}}.
\ee
Using again~\eqref{eq:MFT-factorization}, one has
\be
\avg{(\de_{\hat n_i}\Pt_k)^2\hat\Delta_k}=\avg{(\de_{\hat n_i}\Pt_k)^2}\avg{\hat\Delta_k} =\avg{(\de_{\hat n_i}\Pt_k)^2}\sigma_0[\avg{\hat n_k}]\left(1+{\cal O}(1/N)\right).
\ee
Moreover,
\be
\de_{\hat n_i}\Pt_k=\sum_{r\geq 1} \sum_{\substack{R\cup\{i-k\}\subset S_N\\ |R|=r-1}} \frac{b_r}{|S_N|^r} \prod_{q\in R}\hat n_{k+q}.
\ee
Therefore,
\begin{align}
\avg{(\de_{\hat n_i}\Pt_k)^2}&=\sum_{r,s\geq 1}\frac{b_r b_s}{|S_N|^{r+s}} \sum_{\substack{R,L\subset S_N\setminus\{i-k\}\\ |R|=r-1,\ |L|=s-1}} \avg{\prod_{q\in R\cup L}\hat n_{k+q}}
=\sum_{r,s\geq 1}\frac{b_r b_s}{|S_N|^{r+s}}\sum_{\substack{R,L\subset S_N\setminus\{i-k\}\\ |R|=r-1,\ |L|=s-1}}\avg{\hat n_k}^{|R\cup L|}+{\cal O}(|S_N|/N).
\end{align}
The number of pairs $(R,L)$ with $|R\cup L|=m$ is
\be
\sum_{\substack{R,L\subset S_N\setminus\{a\}\\ |R|=r-1,\ |L|=s-1\\ |R\cup L|=m}}1
=\binom{|S_N|-1}{m}\binom{m}{r-1}\binom{r-1}{r+s-2-m}.
\ee
Thus, defining
\be
A^{(2)}_{m}:=\sum_{r,s\geq 1}\frac{b_r b_s}{|S_N|^{r+s-2}}\binom{|S_N|-1}{m}\binom{m}{r-1} \binom{r-1}{r+s-2-m},
\ee
one obtains
\be
\avg{(\de_{\hat n_i}\Pt_k)^2}=\frac{1}{|S_N|^2}\Big(\sum_{m\geq 0}A^{(2)}_{m}\,\avg{\hat n_k}^{m}\Big).
\ee
Finally, summing over the $|S_N|$ values of $k$ such that $\{i-k\}\in S_N$, and taking the large-scale limit, gives
\be
m^{2}_{(2)}(x;\tau)=c_\lambda\, \sigma_0[\bar n(x;\tau)] \Big(\sum_{m\geq 0}A^{(2)}_{m}\,\bar n(x;\tau)^m\Big), \quad c_\lambda=\lambda^2/\lambda_*^2.
\ee
Both contributions are finite in the scaling regime $\lambda\sim \lambda_*$. The full mass is then obtained as
\begin{align}
m^2(x;\tau)=m^{2}_{(1)}(x;\tau)+m^{2}_{(2)}(x;\tau)=&\sigma_0[\bar n(x;\tau)] \sum_{m\geq 0} c_\lambda\big(A_m^{(1)}+A_m^{(2)}\big) \bar{n}(x;\tau)^m\nn
=&\sum_{m\geq 1} A_m \bar{n}(x;\tau)^m, \quad\text{with}\;
A_m:=2c_\lambda\left[A^{(1)}_{m-1}+A^{(2)}_{m-1}-A^{(1)}_{m-2}-A^{(2)}_{m-2}\right],
\end{align}
where, by convention, $A^{(1)}_{-1}=A^{(2)}_{-1}=0$.\\

For the short-range example in Eq.~\eqref{eq:short-range-dressing} of the main text, using $b_1=2$, $b_2=-8$ and $|S_N|=2$, one obtains the coefficients
\begin{align}
&A_0=0, \qquad A_1=2c_\lambda b_1^2=8 c_\lambda, \qquad A_2=c_\lambda\left(-2b_1^2+2b_1b_2+\frac38 b_2^2\right)=-16c_\lambda,\nn
&A_3=c_\lambda\left(-2b_1b_2-\frac14 b_2^2\right)=16 c_\lambda,\qquad  A_4=-\frac18 c_\lambda b_2^2=-8c_\lambda,
\end{align}
and $A_{m>4}=0$. This gives,
\be
m^2[\bar n]=A_1\bar n + A_2\bar n^2 + A_3\bar n^3 + A_4 \bar n^4=2c_\lambda\sigma_0(\bar n)\big[2-\sigma_0(\bar n)\big].
\ee
Equivalently, this result can be obtained from a direct calculation. Using Eq.~\eqref{eq:short-range-dressing}, one has $\Pt_i^2=\Pt_i$ and therefore
\be
D_0^{-1}a^{(1)}_{i,t} = 2N^2\lambda^2 \left(\avg{\Pt_i^2}_t-\avg{\Pt_i}_t^2\right)= 2c_\lambda\,\sigma_0[\avg{\hat n_i}_t] \big(1-\sigma_0[\avg{\hat n_i}_t] \big).
\ee
For the second contribution in Eq.~\eqref{eq:mass-counting} of the main text,
\be
\de_{\hat n_i}\Pt_{i+1}=1-2\hat n_{i+3}, \qquad
\de_{\hat n_i}\Pt_{i-2}=1-2\hat n_{i-3},
\ee
so that $(\de_{\hat n_i}\Pt_{i+1})^2=(\de_{\hat n_i}\Pt_{i-2})^2=1$. Therefore,
\be
D_0^{-1}a^{(2)}_{i,t} = N^2\lambda^2 \sum_{k:\{i-k\}\in S_N} \avg{(\de_{\hat n_i}\Pt_k)^2\hat\Delta_k}_t= 2c_\lambda\,\sigma_0[\avg{\hat n_i}_t].
\ee
Combining the two contributions, one recovers
\be
D_0^{-1}\big(a^{(1)}_{i,t}+a^{(2)}_{i,t}\big)= 2c_\lambda\sigma_0[\avg{\hat n_i}_t]\big(2-\sigma_0[\avg{\hat n_i}_t]\big).
\ee
\smsubsection{Discussion of the case with $n>2$ replicas}\label{sec:sm-generalization}
In this section, we discuss the generalization of the two-replica result to arbitrary coherence loops. For the $n$-replica loop, we label the legs cyclically, $r=1,\dots,n$, with the $r$th leg corresponding to $i_r\to i_{r+1}$ and $i_{n+1}\equiv i_1$. The It\=o contractions involve pairs of replicas $(r,q)$, with $1\leq r<q\leq n$. The total number of such pairings is $\binom{n}{2}=n(n-1)/2$.\\

We distinguish the adjacent pairings along the loop,
\be
{\cal A}_n:=\{(1,2),(2,3),\dots,(n-1,n),(1,n)\},
\qquad |{\cal A}_n|=n,
\ee
from the non-adjacent pairings, for $n\geq3$,
\be
{\cal N}_n:= \{(r,q):1\leq r<q\leq n\}\setminus{\cal A}_n,
\qquad
|{\cal N}_n|=\binom n2-n=\frac{n(n-3)}2.
\ee
We also introduce the shorthand
\[
\circled{a}_{r;k}:=\circled{a}_{i_r i_{r+1};k},
\qquad
\circled{b}_{r;k}:=\circled{b}_{i_r i_{r+1};k},
\]
and
\[
\circled{c}_{(r,q);k}:=\circled{c}_{i_r,i_{r+1},i_q,i_{q+1};k},
\qquad
\circled{d}_{(r,q);k}:=\circled{d}_{i_r,i_{r+1},i_q,i_{q+1};k}.
\]
Then, neglecting the $\circled{e}$ contribution as discussed above, the equation of motion for the $n$-replica coherence loop can be written as
\begin{align}
\frac{d}{dt}\G_{i_1,\dots,i_n}
=&\sum_{r=1}^n\sum_{k=0}^{N-1} \E\left[\Big(\circled{a}_{r;k}-\circled{b}_{r;k}\Big)\prod_{s\neq r}\langle \hat X_{i_s,i_{s+1}}\rangle_t\right]
\nn
&-\sum_{(r,q)\in{\cal A}_n}\sum_{k=0}^{N-1}\E\left[\Big(\circled{c}_{(r,q);k}+\circled{d}_{(r,q);k}\Big)\prod_{s\notin\{r,q\}}\langle \hat X_{i_s,i_{s+1}}\rangle_t\right]
\nn
&-\sum_{(r,q)\in{\cal N}_n}\sum_{k=0}^{N-1}\E\left[\Big(\circled{c}_{(r,q);k}+\circled{d}_{(r,q);k}\Big)\prod_{s\notin\{r,q\}}\langle \hat X_{i_s,i_{s+1}}\rangle_t\right].
\end{align}
The different terms combine as follows:
\begin{itemize}
\item The QSSEP-like drift terms $\circled{a}_{r;k}$ combine with the adjacent It\=o contractions $\circled{c}_{(r,q);k}$, with $(r,q)\in{\cal A}_n$, to generate the Laplacian acting on each point $i_r$ of the loop. They also produce the contribution $a^{(1)}_{i_r,t}$ in Eq.~\eqref{eq:mass-counting} of the main text at each point $i_r$ of the loop, and the adjacent part of the QSSEP contact source. These adjacent contacts split the original $n$-loop into two loops of sizes $n-1$ and $1$.

\item The QSSEP-like It\=o contractions on non-adjacent pairings, $\circled{c}_{(r,q);k}$ with $(r,q)\in{\cal N}_n$, generate the remaining QSSEP contact sources. These are the terms in which the original $n$-loop is split into two loops of sizes $p$ and $n-p$, with $p>1$.

\item The terms $\circled{b}_{r;k}$ generate the second mass contribution $a^{(2)}_{i_r,t}$ in Eq.~\eqref{eq:mass-counting} of the main text at each point $i_r$ forming the loop. The additional mesoscopic contact terms linear in $\de_{\hat n}\hat P$ which appear in $\circled{b}_{r;k}$ cancel, at leading order, against the corresponding terms coming from the adjacent contractions $\circled{d}_{(r,q);k}$, with $(r,q)\in{\cal A}_n$.

\item The non-adjacent terms $\circled{d}_{(r,q);k}$, with $(r,q)\in{\cal N}_n$, do not generate additional leading contact sources for the coherence loop. Away from contact reductions they are subleading by the power counting discussed above.
\end{itemize}

We do not report the full calculation, since the intermediate expressions are lengthy and do not introduce new structures beyond Sec.~\ref{smsec:multi-replica}, the two-replica calculation, and Ref.~\cite{Hruza2023}. As a final result, one obtains the hydrodynamic equation for the scaling function
\be
g_{n}(x_1,\dots,x_n;\tau):=\lim_{N\to\infty} N^{n-1} \E[\langle \hat X_{i_1,i_2}\rangle_t\dots\langle \hat X_{i_n,i_1}\rangle_t]^c \Big\vert_{\substack{i_k=x_k N \\ t=\tau N^2}},
\ee
that is
\be\Big(\de_\tau -D_0\sum_{r=1}^n\big[\de_{x_r}^2-m^2(x_r;\tau)\big]\Big)g_{n}(x_1,\dots,x_n;\tau)={\cal S}_{n}^{\rm qssep}(x_1,\dots,x_n;\tau).
\ee
Here, $m^2(x_r;\tau)$ is the same mass as in Eq.~\eqref{eq:mass} of the main text, acting on each point of the loop. The QSSEP contact source is
\begin{align}
{\cal S}_{n}^{\rm QSSEP}(x_1,\dots,x_n;\tau):= 2D_0 \sum_{1\leq i<j\leq n}\delta(x_i-x_j) &\big(\de_{x_i}g_{n-p}(\underbrace{x_i,x_{j+1},\dots,x_n,x_1,\dots,x_{i-1}}_{n-p};\tau)\big)\times \nn
&\times\big(\de_{x_j} g_{p}(\underbrace{x_j,x_{i+1},\dots,x_{j-1}}_{p};\tau)\big),
\end{align}
where $g_1(x;\tau)\equiv\bar n(x;\tau)$. The boundary conditions are 
\be
g_n(x_1,\dots,x_n)=\begin{cases}
n_a, n_b; \quad\text{for $n=1$ and $x=0,1$};\\[5pt]
0, \quad\text{for $n\geq 2$ and some $x_i\in\{0,1\}$}.
\end{cases}
\ee
\end{document}